\documentclass[aps,pre,groupedaddress,amsmath,eqsecnum,floatfix]{revtex4}
\usepackage{bm}
\usepackage{graphicx}
\newcommand{\Label}[1]{\label{#1}}                  %% DO NOT show labels
\newcommand{\Bibitem}[1]{\bibitem{#1}}        %% DO NOT show labels
\newcommand\be{\begin{equation}}
\newcommand\ee{\end{equation}}
\newcommand\ba{\begin{eqnarray}}
\newcommand\ea{\end{eqnarray}}
\newcommand{\nn}{\nonumber\\}
\newcommand{\av}[1]{\langle #1\rangle}  %type \av{A} to make <A>

\newcommand{\half}{\textstyle{\frac{1}{2}}}

\newcommand{\bhat}[1]{\hat{{\bf #1}}}

\newcommand{\bg}{{\bf g}}
\newcommand{\bgperp}{{{\bg}_\bot}}
\newcommand{\bn}{{\bf n}}
\newcommand{\bG}{{\bf G}}
\newcommand{\rg}{{g}}
\newcommand{\gperp}{{g_\bot}}
\newcommand{\gpar}{{{g}_\parallel}}
\newcommand{\hgpar}{{\hat{g}_\parallel}}
\newcommand{\bv}{{\bf v}}
\newcommand{\bc}{{\bf c}}
\newcommand{\bw}{{\bf w}}
\newcommand{\br}{{\bf r}}
\newcommand{\bu}{{\bf u}}

\newcommand{\Ref}[1]{(\ref{#1})}

\renewcommand{\theequation}{\arabic{section}.\arabic{equation}}

\begin{document}
%\preprint{}
\title{The Boltzmann equation for driven systems of inelastic soft spheres}
\author{M. H. Ernst$^{1,2}$, E. Trizac$^{3}$ and A. Barrat$^{4}$}
%\email{ernst@phys.uu.nl}
\affiliation{\it $\,^{1}$Instituut voor Theoretische Fysica, Universiteit
Utrecht, Postbus 80.195, 3508 TD Utrecht, The Netherlands, \it
$\,^{2}$Dpt.~de F\'{\i}sica Aplicada I, Universidad Complutense,
E--28040 Madrid, Spain,
{\it$\,^{3}$}Laboratoire de Physique Th\'eorique
et Mod\`eles Statistiques (UMR du CNRS
8626), B\^atiment 100, Universit\'e de Paris-Sud,
91405 Orsay cedex, France
and {\it$\,^{4}$}Laboratoire de Physique
Th\'eorique (UMR du CNRS
8627), B\^atiment 210,
Universit\'e de Paris-Sud, 91405 Orsay Cedex, France}
%\date{June-3-2005}
\date{June 10, 2005}
\begin{abstract}
We study a generic class of inelastic soft sphere models with a
binary collision rate $g^\nu$ that depends on the relative
velocity $g$. This includes previously studied inelastic hard
spheres ($\nu=1$) and inelastic Maxwell molecules ($\nu=0$). We
develop a new asymptotic method for analyzing large deviations
from Gaussian behavior for the velocity distribution function
$f(c)$. The framework is that of the spatially uniform nonlinear
Boltzmann equation and special emphasis is put on the situation
where the system is driven by white noise. Depending on the value
of exponent $\nu$, three different situations are reported. For
$\nu<-2$, the non-equilibrium steady state is a repelling fixed
point of the dynamics. For $\nu>-2$, it becomes an attractive
fixed point, with velocity distributions $f(c)$ having stretched
exponential behavior at large $c$. The corresponding dominant
behavior of $f(c)$ is computed together with sub-leading
corrections. In the marginally stable case $\nu=-2$, the high
energy tail of $f(c)$ is of power law type and the associated
exponents are calculated. Our analytical predictions are
confronted with Monte Carlo simulations, with a remarkably good
agreement.
\end{abstract}
\maketitle

Pacs \#: 45.70, 47.70 Nd; 5.40-a, 81.05 RM.

\renewcommand{\theequation}{1.\arabic{equation}}
\setcounter{section}{0} \setcounter{equation}{0}
\section{Introduction}

The interest in kinetic theory of dissipative systems, such as
granular gases and fluids
\cite{Campbell,Goldhirsch,Brey-kinetic,PTvNE-II,Goldstein-shapiro}
has caused a great revival in the study of the Boltzmann equation
\cite{SpringerII,SpringerI}. Not surprisingly, the introduction of
Maxwell models with their energy independent collision rate
--which simplifies the nonlinear collision term to a convolution
product-- has had a great part in that
\cite{CCG00,BCG00,BNK-PRE00}. In this paper the focus is on the
velocity distribution function (v.d.f.) $F(v,t)$ in spatially
uniform states of inelastic systems, evolving according to
inelastic generalizations of the Boltzmann equation for classical
repulsive power law interactions. For these systems we study the
asymptotic properties of the v.d.f.'s at large times and at large
velocities. This will be done for cases without energy supply,
i.e. {\it freely cooling} systems, as well as for {\it driven}
systems. The latter ones may approach a non-equilibrium steady
state (NESS), and the former ones approach scaling states,
described by scaling or similarity solutions of the nonlinear
Boltzmann equation. Both types of asymptotic states show features
of universality, such as independence of initial states, and
independence of the strength of the energy input, but do depend on
the type of driving device.

The big boom occurred in 2002 after the discovery of an exact
scaling  solution, $f(c) = (2/\pi) (1+c^2)^{-2}$, of the spatially
uniform nonlinear Boltzmann equation (NLBE) for a freely cooling
one-dimensional Maxwell model \cite{Rome1,Rome-Springer}. Here,
the velocity distribution function (v.d.f) has the scaling form
$F(v,t) = v_0^{-d} f(v/v_0)$ where $v_0(t)$ is the r.m.s velocity
and $d$ is the number of spatial dimensions. Subsequent analysis
\cite{BNK-JPA02,BNK-Springer,EB-EPL,EB-Rap} has shown that the
NLBE for freely cooling inelastic Maxwell models in $d$ dimensions
has a scaling solution with a power law tail $f(c)\sim 1/c^s$. The
power law exponent $s$ can be calculated from a transcendental
equation, and depends on the dimensionality $d$ of the system and
on the coefficient of restitution $\alpha$, where $1-\alpha^2$
measures the fractional energy loss in an inelastic collision. The
proof that the v.d.f. $F(v,t)$ for general initial values $F(v,0)$
approaches for large $t$ this scaling solution has been given in
\cite{AB+CC-proof}. The scaling and  NESS solutions mentioned
above have the remarkable property of possessing a power law tail,
$f(c)\sim 1/c^s$, which is highly overpopulated at large
velocities as compared to a Maxwellian v.d.f $\sim \exp(-c^2)$.
Solutions with overpopulated tails of stretched exponential form,
$f(c)\sim \exp(-\beta c^b)$ with $0<b<2$, have been studied before
both analytically and by numerical simulations (both Molecular
Dynamics and Monte Carlo) for inelastic hard spheres
\cite{Esipov,vNE-granmat,Brey-scaling,MS00,EB-Rap,PTvNE-II,BBRTvW01,Gamba1,Gamba2,Gamba3},
as well as for inelastic soft spheres
\cite{EB-Springer,BNK-Springer}. Such states appear not only in
freely cooling systems but also in driven systems. These systems
are in principle more interesting because the collisional loss of
energy may be compensated by spatially uniform heating devices.
They may drive the system into a NESS, with possibilities of
experimental verification.

Many different types of heating devices have been invented:
\begin{enumerate}
\itemsep=0pt
\item[\em i)] white noise, which adds random velocity increments to the
particles
\cite{WM96,CCG00,BNK-Springer,BNK-JPA02,MS00,vNE-granmat,preIgn,EB-Springer,EB-Rap,PTvNE-II}
\item[\em ii)] a deterministic friction force
${\bf a} = \gamma_0 \hat{\bf v} |v|^\theta$  with $\theta \geq 0$
and $\hat{\bf v} = {\bf v}/v$ (nonlinear friction model). For
$\theta=1$ it gives the Gaussian thermostat equivalent to a linear
rescaling of the velocities, and for $\theta=0$ it gives the
`gravity' thermostat, which models something like gliding friction
\cite{MS00}. These thermostats have been used in theoretical,
molecular dynamics and Monte Carlo studies
\cite{Brey-scaling,MS00,Sinai,Soto,BBRTvW01,EB-Rap,Evans-book} to
add ($\gamma_0>0$) or to remove ($\gamma_0<0$) energy from the
system.
\item[\em iii)] a heating device, recently proposed in Ref.
\cite{EBN-Machta}, that feeds energy into the system  in the
ultra high energy tail of the v.d.f and forces the system
into a steady state with a power law tail. We shall come back to this
model at the end of section \ref{sec:VI}.
\item[\em iv)] a heat bath consisting of an ideal gas kept in a thermal
equilibrium state \cite{Piasecki}, or a more exotic device that
maintains the bath in some non-equilibrium steady state in which
the v.d.f. is, for instance, described by a L\'evy distribution,
having power law tails \cite{Barkay}.
\end{enumerate}

Over-populated tails, corresponding to large deviations from the
Maxwellian v.d.f, have also been observed in many experimental
studies of driven granular gases and fluids \cite{GM-exp}. The
heating devices used there are difficult to model in a simple
kinetic theory context \cite{EPJEBT}.

The existence of such tails is remarkable because the standard
NLBE for classical gases and fluids with conservative interactions
\cite{Chapman,Resibois,Cercignani-book,Uhl+Fo63} provides the
basic notion of rapid relaxation of a general initial distribution
$F(v,0)$ towards a steady state described by a Maxwellian v.d.f.
Here the steady state solutions satisfy the detailed balance
relation $F({\bf v}'_1,\infty) F({\bf v}'_2,\infty) = F({\bf
v}_1,\infty) F({\bf v}_2,\infty)$ between pre-collision velocities
$({\bf v}_1,{\bf v}_2)$ and post-collision ones $({\bf v}'_1,{\bf
v}'_2)$. The detailed balance relation in combination with the
$H$-theorem forces the steady state solution to be a Maxwellian.
The basic reason for the large deviations from Maxwellian behavior
is the violation of detailed balance in dissipative collisions,
together with the breakdown of the $H$-theorem \cite{EB-EPL}.

The goal of this paper is to develop a new asymptotic method for
analyzing the large deviations from Maxwellian behavior in the
high energy tails of the v.d.f. observed in NESS solutions of the
nonlinear Boltzmann equation for dissipative systems, driven by an
energy source. The class of $\nu$ models studied are the {\em
inelastic soft spheres} \cite{EB-Springer,EBN-Machta} with a
binary collision rate, scaling like $g^\nu$, where $g$ is the
relative speed. It includes the previously studied inelastic hard
spheres ($\nu=1$) and inelastic Maxwell molecules ($\nu=0$).

The method can be applied to all driving devices listed above.
Depending on the device used, the models have in general either
{\em i)} a {\em stable} NESS, i.e attracting fixed point solution,
for $b>0$ or $\nu>\nu_c$ with stretched exponential tails
$f(c)\sim \exp(-\beta c^b)$ and higher asymptotic corrections,
like  $f(c)\sim c^\chi \exp(-\beta c^b + \beta' c^{b'})$, {\em
ii)} a {\em marginally stable} NESS for a threshold model with
$b=0$ or $\nu=\nu_c$ with power law tails, and {\em iii)} an {\em
unstable} NESS, i.e. a repelling fixed point solution for $b<0$ or
$ \nu<\nu_c$. For free cooling and Gaussian thermostats $b=\nu$,
for white noise driving $b=1+\half\nu$ and for nonlinear friction
models $b=\nu+1 - \theta$. The preliminary results have been
published in Ref. \cite{Pucon}. The present paper focusses on the
mathematical method, and on the simple application of white noise
driving (i), and on the ultra high energy source (iii).
  The remaining results will be published elsewhere
\cite{ETB-II}.

The crux of the new asymptotic analysis is the construction in
Section \ref{sec:II} of a linearized collision operator, whose
eigenfunctions are powers of the velocity $c$, and whose
eigenvalues determine the power law exponent in the tail of the
distributions. The Fourier transform method used in Refs.
\cite{BNK-JPA02,EB-EPL} for determining power law tails can only
be applied to Maxwell models ($\nu=0$). In Section \ref{sec:III}
we study the spectral properties of the aforementioned operator,
with details given in appendices \ref{app:A} and \ref{app:B}.
These two sections plus appendices can be considered as the
generalization to inelastic soft spheres of the mathematical
theory for inelastic Maxwell molecules in Refs.
\cite{CCG00,BCG00}. In Section \ref{sec:IV} we derive the energy
balance equation for the white noise driven case, determine the
stability of the fixed point solutions, and derive the integral
equation for the scaling form $f(c)$. In Section \ref{sec:V} we
present the high energy tails of exponential type, and in Section
\ref{sec:VI} those of power law type,  where also the ultra
high energy source is discussed.  The results are supported
by Direct Monte Carlo Simulations (DSMC scheme \cite{Bird}). We
end in Section \ref{sec:VII} with conclusions and perspectives.

\renewcommand{\theequation}{2.\arabic{equation}}
\setcounter{section}{1} \setcounter{equation}{0}
\section{Basics of inelastic scattering models}
\label{sec:II}

The nonlinear Boltzmann equation for dissipative interactions in a
spatially  freely evolving state can be put in a broader
perspective, that covers both elastic and inelastic collisions, as
well as interactions where the scattering of particles is
described by deterministic (conservative or dissipative) forces or
by stochastic ones. To do so it is convenient to interpret the
Boltzmann equation as a stochastic process, similar to the
presentations in the classical articles of Waldmann
\cite{Waldm58}, Uhlenbeck and Ford \cite{Uhl+Fo63}, or in
Ulam's stochastic model \cite{Ulam}. The last one shows the basics
of the approach of a one-dimensional gas of elastic particles
towards a Maxwellian distribution.

We consider a spatially homogeneous fluid or gas of elastic or
inelastic particles in $d$ dimensions, described by an {\it
isotropic} velocity distribution function, $F(v,t)= F(|\bv|,t)$,
and evolving  according to the nonlinear Boltzmann equation,
\ba \Label{BE-scatt}
&\partial_t F(v,t)=  I(v|F)\equiv \int  d\bw d\bv^\prime
d\bw^\prime\int d\bn \left[W(\bv,\bw|\bv^\prime,\bw^\prime;\bn)
\right.& \nn &\left.\times F(v^\prime,t)F(w^\prime,t) -
W(\bv^\prime,\bw^\prime|\bv,\bw;\bn)F(v,t)F(w,t) \right],&
\ea
where the binary collisions are described through a transition
probability per unit time, $W(\bv^\prime,\bw^\prime|\bv,\bw;\bn)$.
The loss term $I_{{\rm loss}}$ and the gain term $I_{{\rm gain}}$
represent respectively the contributions from the direct collisions
$(\bv,\bw) \to (\bv^\prime,\bw^\prime)$, and from the restituting
collisions, $(\bv^\prime,\bw^\prime) \to (\bv ,\bw )$. Here the
direct and restituting velocities have been parameterized in terms
of the incoming velocities $(\bv,\bw)$, and an impact (unit) vector
$\bn$, that is chosen  on the surface of a unit sphere with a
probability proportional to the collision frequency
$K(\gperp,\gpar)$. It depends only on the length $\rg_\bot =
|\bgperp|$ and $|\gpar|$ of the vector arguments. We further use
the notations $a_\parallel =
\bf{a}\cdot \bn $ and $ {\bf{a}}_\bot = {\bf{a}} -$ $
{a_\parallel}{\bn}$, and $\bhat{a} ={\bf{a}} / a$ is a unit vector.

The transition probability for the inelastic collisions,
$(\bv,\bw) \to (\bv^\prime,\bw^\prime)$, is given  by
\cite{EB-Springer},
\be \Label{W-gen}
W(\bv^\prime,\bw^\prime|\bv,\bw;\bn) = {K}(\gperp,\gpar)
\delta^{(d)}(\bG^\prime -\bG)\delta^{(d-1)}(\bg^\prime_\bot -\bg_\bot)
\delta({\rg^\prime}_\parallel + \alpha \gpar),
\ee
where $\bG =\half(\bv+\bw)$, $\bg = \bv - \bw$, and $\alpha $
obeys $0 \leq \alpha <1$. In the subsequent analysis, we refrain
from explicitly indicating the dimension of the argument for the
Dirac functions. The value $\alpha = 1 $ corresponds to the
elastic case, and $\alpha=0$ to the totally inelastic case.
The elastic case, possibly driven by by source(s)
and/or sink(s), will not be considered here. On the other hand,
the quasi-elastic limit $(\alpha \to 1)$ possibly coupled to
large-$c$ limits is certainly of interest for inelastic gases
\cite{BBRTvW01,Santos+E}.

The transition rate \Ref{W-gen} implies that the
scattering laws are in all models the {\it same} as for inelastic
{\it hard spheres}, i.e. as in an inelastic collision of two
perfectly smooth spheres without rotational degrees of freedom,
where  $\bG$ and $\bg_\perp$ are conserved, and
$g^\prime_\parallel = - \alpha g_\parallel$ is reflected and
reduced in size by a factor $\alpha$. This implies for the {\it
direct} collisions, $ (\bv,\bw) \rightarrow (\bv^*,\bw^*)$,
\ba \Label{coll}
\bv^\ast &\equiv & \bv^\ast(\alpha)= \bv -\half (1+\alpha)\gpar  \bn
\nn \bw^\ast &\equiv & \bw^\ast(\alpha)= \bw +\half (1+\alpha)\gpar \bn.
\ea
The corresponding energy loss in such a collision is, $ \Delta E =
\half( v^2 +w^2 - v^{*2} -w^{*2})= pq|\gpar|^2$, where $q = 1-p =
\half (1-\alpha)$ measures the degree of inelasticity. The restituting
velocities, describing $(\bv^{**},\bw^{**}) \to (\bv,\bw) $, are
given by the inverse transformation of \Ref{coll}, $
\bv^{**}= \bv^*(1/\alpha)$ and $\bw^{**}=
\bw^*(1/\alpha)$.

A second aspect of the interaction dynamics is the collision rate,
$K \sim g \sigma (g) $. In {\it elastic} cases the rate is
proportional to the differential scattering cross-section
$\sigma$, which depends in general on the relative speed $g =
|\bg|$, and on the scattering angle, which is related to the
impact vector $\bn $ \cite{Chapman}. For elastic hard spheres
$\sigma (g) = const$, and $K \sim g$, and for Maxwell molecules $K
= const$.  For repulsive {\it power law} potentials, $V(r) \sim
r^{-n}$, referred to as soft elastic spheres, the collision
frequency scales as $ K \sim g \sigma(g)\sim g^\nu$, and the
exponent $\nu$ is related to the exponent $n$ in the power law
potential through $\nu = 1-2(d-1)/n$ \cite{ME-PhysRep}. For
positive $n$ the exponent $\nu$ is restricted to  $ -\infty <\nu
\leq 1$. As an inelastic generalization of soft elastic spheres we
propose systems with a collision frequency that scales as
$K(\gperp,\gpar) \sim \rg^\nu |\hat{\rg}_\parallel|^\sigma =
{\rg}^{\nu-\sigma} |\gpar |^\sigma $, where  the exponent $\nu$
parameterizes the energy dependence and $\sigma$  the dependence
on the angle of incidence $\theta$, defined through $\hgpar =
\bhat{g} \cdot \bn = \cos \theta$. After inserting $K$ in
\Ref{BE-scatt} two velocity integrations can be carried out using
the delta functions in \Ref{W-gen}, and the spatially homogeneous
Boltzmann equation reduces to \cite{EB-Springer},
\be\Label{BE}
\partial_t F(v)= I (v|F)\equiv \int_\bn  \int d\bw
\left[ \frac{1}{\alpha} K(\gperp,\textstyle{\frac{\gpar}{\alpha}})
F(v^{**}) F(w^{**}) - K(\gperp,\gpar)F(v) F(w) \right].
\ee
Here $\int_\bn (\cdots) = (1/\Omega_d) \int d\bn(\cdots)$ is an
angular average over the surface area $\Omega_d = 2 \pi^{d/2}/
\Gamma(\half d)$ of a $d$-dimensional unit sphere. We have
absorbed constant factors in the time scale. In the one-dimensional
case $\int_\bn \to 1$ and $ \bg \to
\gpar, \bgperp
\to 0 $.

To clarify the meaning of the exponent $\sigma$ in the collision
rate,  we consider the impact parameter $b$, defined as $b
=|\bhat{g}\times \bn| =\sin
\theta$ and $db = \cos\theta d\theta $. The distribution ${\cal
P}(b) $ of impact parameters can be obtained from,
\be \Label{P-b}
\int d\bn |\hgpar|^\sigma \sim \int^{\pi/2}_0 d\theta (\sin
\theta)^{d-2}|\cos\theta |^\sigma
 \sim \int^1_0 db b^{d-2} (1-b^2)^{(\sigma -1)/2} \equiv \int^1_0
db b^{d-2}{\cal P}(b).
\ee
This shows that ${\cal P}(b) \sim (1-b^2)^{(\sigma -1)/2}$ for $
\sigma <1$ is {\it biased} towards {\it grazing} collisions, and
for $\sigma >1$ towards {\it head on } collisions. The cases with
$\sigma \neq 1$, correspond to pre-collision velocity correlations
between the colliding particles. This is a violation of
Boltzmann's basic postulate of molecular chaos. For dimensions
$d>1$ {\it molecular chaos} requires $\sigma =1$ or $K \sim
{\rg}^\nu |cos \theta|$, corresponding to a uniform distribution
${\cal P}(b)=1$. For {\it mathematical} convenience it looks
attractive to set $\sigma=\nu$, and have the simple collision
frequency $K=|{\rg}_\parallel|^\nu$. Then the Boltzmann equation
takes the simple form \cite{EB-Springer},
\be \Label{BE-nu}
\partial_t F(v)=   \int_\bn  \int d\bw |\gpar|^\nu
\left[{\alpha}^{-\nu-1} F(v^{**}) F(w^{**}) - F(v) F(w) \right].
\ee
Equation \Ref{BE-nu} becomes quite simple for inelastic Maxwell
models ($\nu=0$), where all angular dependence in the collision
frequency is absent \cite{BNK-JPA02,EB-EPL,EB-Springer}. The
Boltzmann equation with $K \sim |\cos \theta|$ for the Maxwell
model with molecular chaos, has been studied in
Refs. \cite{BCG00,CCG00,EB-EPL}.

Regarding the subject of interest in this paper, i.e. high energy
tails of $F(v,t)$ in {\it spatially homogeneous} systems, the
differences between the classes of models with different values of
$ \sigma$ are expected to be mostly of a qualitative nature
\cite{EB-EPL}, at least for $\nu \geq 0$. As it turns out the
shape of the distribution function, in particular its high energy
tail, depends in a sensitive way
\cite{vNE-granmat,EB-Springer,MS00,EB-Rap} on the energy
dependence of the collision rate at large energies, and not that
much on the scattering angle. In the present paper the Boltzmann
equation \Ref{BE} is studied for general exponents $\nu$ and
$\sigma$, where cases with $\sigma \neq 1$ do not introduce any
additional complications. Most Monte Carlo simulations (see Ref.
\cite{Rome1}) and most analytic studies
\cite{EB-Springer,BNK-Springer,BNK-JPA02} have been carried out
for models with biased distributions, ${\cal P}(b) \sim
(1-b^2)^{(\sigma - 1)/2}$, and only a few studies exist for
molecular chaos model with $\sigma =1 $ or ${\cal P}(b) =1$
\cite{BCG00,BC02-JSP,EB-EPL}.

Equation \Ref{BE} is a generalization of the elastic Boltzmann
equation to a general class of inelastic models with collision
rate $K \sim g^\nu |\cos \theta|^\sigma $, which include all
models presently known in the literature. For $\nu=1 \: (n
=\infty) $ one has inelastic hard spheres and $\nu=0$ corresponds
to  the softer inelastic Maxwell models $(n=2(d-1))$. Negative
values of $\nu$, as $n$ decreases further, are also possible. They
would correspond to still softer interactions. Inelastic models
with $\nu \geq 1$ have also been studied. In the Boltzmann
equation with elastic scattering laws there also exists a
stochastic scattering model, the {\it Very Hard Particle} model
with $\nu =2$, for which the two-dimensional homogeneous NLBE has
been solved exactly as an initial value problem \cite{ME-PhysRep}.
So, there is no a priori mathematical reason to impose
restrictions on the values of $\nu$. Regarding the $\sigma$
exponent  we require that the mean collision rate remains bounded.
This implies that the angular average appearing in the loss term
of \Ref{BE} should converge. This imposes the restriction $\sigma
> -1$ on the models in \Ref{BE} (see \Ref{A5} in Appendix A).
Velocities and time have been dimensionalized in terms of the
width and the mean free time of the initial distribution.
Moreover, the Boltzmann equation obeys conservation of mass and
total momentum, but the average kinetic energy or granular
temperature, $T\sim \av{ v^2}_t=v_0^2(t)$, decreases in time on
account of the dissipative collisions, i.e. $ \int d\bv (1,\: \bv
,\: v^2) F(v,t)= (1,\:0,\:\half d v_0^2(t))$, where $v_0(t)$ is
the r.m.s. velocity or width of $F(v,t)$.

To reach a spatially homogeneous steady state, energy has to be
supplied homogeneously in space. This will be done here by
connecting the system to a thermostat or heating device, as
discussed in the introduction. For instance, a {\it negative}
friction force  may be used as a heating device to compensate for
the dissipational losses of energy. Complex fluids (e.g. granular
matter) subject to such forces can be described --in between
collisions-- by the microscopic equations of motion for the
particles, $\dot{\br}_i =\bv_i$, and $\dot{\bv}_i = {\bf a}_i +
\tilde{{\bf \xi}}_i$ $(i=1,2, \cdots)$, where ${\bf a}_i$ is a
possible conservative or friction force per unit mass, and
$\tilde{{\bf \xi}}_i$ a random force. If the system is driven by
Gaussian white noise (WN), the forcing can be represented by
adding a diffusion term \cite{WM96}, $-D {\bf
\partial}^2 F$,  to  the Boltzmann equation.
A negative friction force can be included into the Boltzmann
equation by adding a force term $(\partial /\partial \bv) ({\bf
a}F)$. The Boltzmann equation for system driven in this manner
reads,
 \be \Label{BE-driven}
\partial_t F(\bv)+ {\cal F}F = I(v|F),
\ee
where the source term takes the form,
\be \Label{source}
{\cal F} F = {\partial} \cdot ({\bf a}F) - D \partial^2 F =
\gamma_0 {\partial} \cdot (\hat{\bv} v^\theta  F) - D {\partial}^2
F.
\ee
Here $ \partial \equiv \partial /\partial \bv$ is a gradient in
$\bv-$space, and $\gamma_0$ and $D$ are positive constants. When
energy is supplied at a constant rate, driven dissipative systems
can reach a NESS. As expected, the elastic case,
driven by white noise, does not reach a steady state\cite{Gamba1}.

In the {\it free cooling} case the inelastic interactions decrease
the kinetic energy of the colliding particles, the r.m.s. velocity
$v_0(t)$ decreases, and the velocity distribution $F(v)\to
\delta(\bv)$ as $t \to \infty$. This scenario is supported by the
nonlinear Boltzmann equation \Ref{BE}, because the distribution
$\delta (\bv)$ is invariant under the collision dynamics,
\be \Label{29}
I(v|\delta)=0,
\ee
as shown in Appendix A1. In general $F (v)$ will be
non-Maxwellian, and its shape will depend on the model parameters
$\nu$ and $\sigma$. The effects of the energy dependence of the
collision rate $K \sim g^\nu$ on the high energy tail of the
scaling form $f(c)$ can be understood intuitively as follows. A
tail particle with $v \gg v_0$ has a collision rate  $K \sim |\bv
-\bw|^\nu \simeq v^\nu$. The smaller $\nu$, the smaller the value
of $K$ in the tail, and the slower the tail particles loose and
redistribute their energy over the thermal range $ v \lesssim
v_0$, and the slower the tail shrinks as a function of $v$, when
compared to the bulk velocities. This leads to an increase in
over-population of the tail. The reverse scenario applies when
increasing $\nu$, leading to a decrease of over-population.

Intuitive pictures of the effects of the heating devices can also
be developed. The linear Gaussian thermostat with ${\bf a} =
\gamma_0 \bv$ simply produces a linear rescaling of the
velocities, and has the nice property of supplying energy to or
subtracting energy from the system without changing the scaling
form $f(c)$. The nonlinear friction force ${\bf a} = \gamma_0
\hat{\bv} v^\theta $ produces a nonlinear rescaling, $A(v)\, \bv$.
For $\theta >1$ it represents a heating device that puts
selectively energy into the tail particles, thus leading to an
increase of over-population of the tail. For $\theta <1$ the
reverse scenario applies. The white noise forcing on the other
hand adds randomly velocity increments $\Delta \bw$, drawn from a
uniform $d-$dimensional distribution, to thermal and tail
particles. So, it is more efficient in redistributing and
randomizing velocities of tail particles, especially at smaller
$\nu-$values.

If the collision frequency $K \propto {\rg}^\nu $ gets  smaller
(c.q. larger) at large relative velocities, then the tail
distribution shrinks slower (c.q. faster) than a Maxwellian,
resulting in an overpopulated (c.q. underpopulated) high energy
tail, described by a stretched  (c.q. compressed) Gaussian, $F(v)
\sim \exp[-\beta v^b]$ with $0 <b \leq 2$ \, (c.q. $ b \geq 2$).
The limit as $b \to 0^+$ corresponds to  heavily overpopulated
power law tails. Similar scenario's presumably occur also when
dissipative systems are coupled to an energy source, and are
slowly heating up or reaching a non-equilibrium steady state
(NESS).

To study the behavior of the v.d.f. with a shrinking width we
consider the distribution function $f(c)$, rescaled by the r.m.s.
velocity $v_0$, i.e. $F(v,t) = v_0^{-d} f(v/v_0)$. Suppose that
for {\it large} $c = v/v_0$ the velocity distribution can be
separated into two parts, $f(c) = f_0(c) + h(c)$, where $h(c)$ is
the {\it singular} tail part, and $f_0(c)$ the
presumably {\it regular} bulk part \cite{footnote}. The
tail part $h(c)$ may be exponentially bounded $\sim \exp[-\beta
c^b] $ with $0<b<2$, or of power law type. In the bulk part
$f_0(c)$ the variable $c$ is effectively restricted to {\it bulk
values} in the thermal range $v \lesssim v_0$ or  $c \lesssim 1$.

To obtain an asymptotic expansion of $f(c)$ we consider the ansatz,
\be \Label{30}
f(c) =\delta(\bc) + h(c),
\ee
and we linearize the collision term $I(c|f)$ around the delta
function. This defines the linearized Boltzmann collision operator
$\Lambda$ as,
\be \Label{31}
I(c|\delta + h) = - \Lambda h(c) + {\cal O}(h^2),
\ee
where Eq.\Ref{29} has been used. Here we restrict ourselves to
isotropic functions $h(c)$, depending only on $c =|\bc|$. As will
be shown in the next section the eigenfunctions of $\Lambda$ decay
like powers $c^{-s}$. Consequently they are very suitable for
describing power law tails, $f(c) \sim c^{-s}$.

In case the high energy behavior of the v.d.f.'s are stretched
Gaussians $\sim \exp[-\beta c^b]$, the above eigenfunctions are no
longer useful, and we have developed a method to determine an
asymptotic expansion of the form,
\be \Label{51a}
\ln f(c) \sim - \beta c^b + \beta' c^{b'} +\chi \ln c + \cdots,
\ee
as will be discussed later.
\\

A comment seems in order here. The delta function
in \Ref{31} does {\it not} mean that the thermal bulk part
$f_0(c)$ ( with $c > 1)$  is singular, as explained in the
paragraph above \Ref{31}. In fact, it is irrelevant for our
arguments of extracting  the dominant large-$c$ singularity
whether the thermal part may or may not contain any singularities.
The separation of $f(c)$ into a 'regular' part $f_0(c)$ and a
singular part $h(c)$ is completely analogous to the method used in
Refs.\cite{BNK-Springer,BNK-JPA02,EB-EPL,EB-Rap} for predicting
the power law tail for Maxwell models using the Fourier transform
method. Let the Fourier transform of $f(|c|)$ be ${\cal F}_k
f(|c|) = \hat{f}(k)$. Its small-$k$ behavior takes the form
$\hat{f}(k) = 1+ k^2 f_2 + k^4 f_4 + \cdots +Ak^\alpha+ \cdots $
with $\alpha \neq$ even integer.  Here the series expansion in
powers of $k^2=|k|^2$ represents the regular part, ${\cal F}_k
f_0(c) = \hat{f}_0(k)= 1+k^2 f_2 +k^4 f_4+\cdots $, whereas
$Ak^\alpha$ represents the dominant the small $k^2$-singularity of
$\hat{h}(k)$. Subsequent Fourier inversion of the small-$k$
behavior of $ \hat{f}(k)$ shows that the regular bulk part
${f_0}(c)$ can be viewed in zeroth order of approximation as
$\delta(c)$. For $c > 1$ only the tail contribution of $h(c) =
f(c)-\delta(c)$ survives. The dominant small-$k^2$ singularity of
$\hat{h}(k) $ corresponds to the dominant large-$c$ singularity of
$h(c)$.

\renewcommand{\theequation}{3.\arabic{equation}}
\setcounter{section}{2} \setcounter{equation}{0}

\section{Spectral properties of collision operators}
\label{sec:III}

The first part of this section is devoted to a study of the
eigenvalue problem of the linearized Boltzmann collision operator
$\Lambda$. This is most conveniently done by first constructing the
adjoint operator, $\Lambda^\dagger$, and determining its eigenvalues
and eigenfunctions, also referred to as (left) L-eigenfunctions of
$\Lambda$. The adjoint is defined through
\be\Label{32}
\langle k | \Lambda h \rangle = \langle h | \Lambda^\dagger k \rangle ,
\ee
where the inner product is $\langle k | h \rangle =\int d\bc k(c)
h(c)$, and $k(c)$ and $h(c)$ are isotropic. This can be done by
using the relation,
\be\Label{33}
\langle k | I(F) \rangle = \frac{1}{2}
\int_{\bn} \int d\bc d\bw K(\gperp,\gpar) F(c) F(w)
\left[
k(c^*)+k(w^*) - k(c)-k(w)
\right]
\ee
with $\bg = \bc -\bw$, as follows from \Ref{BE}. To facilitate the
analysis we express $\bc^*$, $\bw^*$ in Eq. \Ref{coll} in terms of
$\bc_\bot$, $\bw_\bot$, $c_\parallel$, $w_\parallel$ with
coefficients $p=\frac{1}{2}(1+\alpha)$ and
$q=\frac{1}{2}(1-\alpha)$ (see Appendix A1). Substitution of
\Ref{30} into \Ref{33}, and linearization yields
\ba \Label{34}
&\langle k | I(\delta+h)\rangle \simeq - \langle k | \Lambda h
\rangle = - \langle h | \Lambda^\dagger k \rangle& \nonumber \\
& = -\int_{\bn} \int d\bc K(c_\bot,c_\parallel) h(c) \left[
k(c)+k(0)-k(|\bc_\bot+\bn q c_\parallel|) -k(|\bn p c_\parallel|)
\right]&,
\ea
where $\delta(\bw)$ has been integrated out. As $k(\bc)=k(|\bc|)$
is isotropic, the adjoint becomes
\be\Label{35}
\Lambda^\dagger k(c)= \int_\bn K(c_\bot,c_\parallel)
\left[
k(c)+k(0)-k(c|\hat{c}_\bot+\bn q \hat{c}_\parallel|) - k(p c
|\hat{c}_\parallel|)
\right]
\ee
with $K(c_\bot,c_\parallel)=c^{\nu-\sigma} |c_\parallel|^\sigma$.
In one dimension, where $\bc_\bot=0$ and $\int_\bn \to 1$, the
adjoint takes the simple form
\be\Label{36}
\Lambda^\dagger k(c)= |c|^\nu \left[
k(|c|)+k(0)-k(q|c|)-k(p|c|)
\right]\ .
\ee
Inspection shows that insertion of $k(c)=|c|^s$ generates a new
power of $|c|$, which solves our eigenvalue
problem
\be\Label{37}
\Lambda^\dagger |c|^s =
|c|^{s+\nu} \left[ 1+\delta_{s,0} - q^s -p^s \def \lambda_s
|c|^{s+\nu}
\right]\ ,
\ee
where $\lambda_s = 1+\delta_{s,0} - q^s -p^s \def \lambda_s
|c|^{s+\nu}$ is the eigenvalue, valid for $s \ge 0$. Strictly
speaking, $|c|^s$ is not an eigenfunction of $\Lambda^\dagger$ but
Eq. \Ref{37}  shows that the family of power law functions is
stable under the action of $\Lambda^\dagger$. In the following, we
will employ the terminology ``eigenvalue'' and ``eigenfunction''
with a similar slight abuse of vocabulary. The same argument shows
with the help of the homogeneity relation
$K(c_\bot,c_\parallel)=c^\nu K(\hat{c}_\bot,\hat{c}_\parallel)$,
that $k(c)=c^s$ with $s\ge 0$ is also an eigenfunction of
$\Lambda^\dagger$ for dimensions $d > 1$. The eigenvalue follows
as
\be\Label{38}
\lambda_s(\sigma) = \int_\bn |\hat{c}_\parallel|^\sigma
\left[
1+\delta_{s,0} - |\hat{c}_\bot +\bn q \hat{c}_\parallel|^s - p^s
|\hat{c}_\parallel|^s
\right] \ .
\ee

For fixed $\hat{\bc}$ the integrand of this $(d-1)$ dimensional
integral is a function of $\hat{c}_\parallel=\hat{\bc}\cdot
\bn=\cos
\theta$. It reduces to a single integral over a polar angle
($0 < \theta < \pi$) for $d\ge 3$, and for $d=2$ to an integral
over an azimuthal angle ($0 < \theta < 2\pi$). The integral is
evaluated in Appendix A2, with the result valid for $s\ge 0$,
\be \Label{39}
\lambda_s (\sigma)= \beta_\sigma \left\{ 1+\delta_{s,0} -
 _2\!F_1\,\left(\textstyle{-\frac{s}{2},
\frac{\sigma+1}{2};\frac{\sigma +d}{2}} \,|\,1-q^2\right)\right\} -p^s
\beta_{s+\sigma}.
\ee
Here $_2F_1(a,b;c|z)$ is a hyper-geometric function and $\beta_s =
\int_\bn |\hat{a}\cdot \bn|^s$ is an average of $|\cos \theta|^s$
over a solid angle, given in \Ref{A5}. It is well defined for
$\sigma >-1$.

So far we have constructed in \Ref{37}-\Ref{38} the eigenvalues
$\lambda_s(\sigma)$ and corresponding eigenfunctions $k_s(c)=c^s$
of the adjoint $\Lambda^\dagger$, which are the (left)
L-eigenfunctions of $\Lambda$. To determine the (right)
R-eigenfunctions $h_s(c)$, corresponding to $\lambda_s(\sigma)$,
we need to construct $\Lambda$ in \Ref{31} explicitly. The steps
are rather technical, and are carried out in Appendix A3. However,
the procedure is quite similar to the steps from \Ref{35} to
\Ref{38}. Once $\Lambda$ is constructed, inspection shows again
that $\Lambda c^r=\mu_r c^{r+\nu}$, where the new eigenvalue
$\mu_r$ is different from $\lambda_s$. Next, $r$ in $\mu_r$ is
chosen such that $\mu_r=\lambda_s$. This yields $r=-s-d-\nu$.

The most important spectral properties for our purpose are:\\
\noindent (i) For $s > 0$ the eigenvalues and R- and L-
eigenfunctions are \cite{Pucon} (with details derived in
\Ref{A20} and \Ref{37}-\Ref{38}),
\ba \Label{313}
\Lambda c^{-s-d-\nu} &=& \lambda_s(\sigma)c^{-s-d}
 \nn    \Lambda^\dagger  c^s  &=& \lambda_s(\sigma)
c^{s+\nu} \ .
\ea
For $s=0$ one has the stationary eigenfunctions, which are
invariant under collisions,
\be \Label{314}
\Lambda \delta(\bc) = 0 \qquad \mbox{and} \qquad \Lambda^\dagger \cdot 1 = 0
\qquad (\lambda_0(\sigma)=0) \ .
\ee
R- and L-eigenfunctions are different because $\Lambda$ is not
self-adjoint. There is in fact among the eigenfunctions in
\Ref{313} another, less trivial, stationary R-eigenfunction,
$c^{-a^*-d-\nu}$, where $s=a^*$ is the root of
$\lambda_s(\sigma)=0$ (see  Fig.\ref{fig2.1}).

\noindent (ii) The spectrum exists for all $s \ge 0$. It is
continuous for $s
> 0$, and $s=0$ is an isolated point of the spectrum. At $s=0$ one
has $_2F_1(0,b;c|z)=1$ and consequently
\be \Label{310}
\lambda_0(\sigma)=0 \qquad \mbox{and}\qquad
\lim_{s\to 0^+} \lambda_s(\sigma)=-\beta_\sigma \ .
\ee
The zero eigenvalue expresses the conservation of mass in binary
collisions. In the elastic limit $\lambda_2(\sigma, \alpha=1)=0$,
which expresses the conservation of energy (see Fig.\ref{fig2.1})
.

\noindent (iii)  As $\beta_{s+\sigma}$, given in \Ref{A5}, and
$\lambda^{(0)}_s(\sigma)$, given in \Ref{A6}, are monotonically
decreasing with $s$, $\lambda_s(\sigma)$ is {\em monotonically
increasing} with $s$, and approaches
\be \Label{311}
\lim_{s\to \infty} \lambda_s(\sigma) = \beta_\sigma.
\ee

\noindent (iv) $\lambda_s(\sigma)$ is {\em independent} of $\nu$,
i.e. the energy-dependence of the collision rate $K \sim g^\nu
|\hat{g}_\parallel|^\sigma$ does not affect the eigenvalue. So, it
is the same for inelastic Maxwell molecules, hard spheres, very
hard particles, and very weakly interacting particles, as modeled
by Eqs. \Ref{BE-scatt} - \Ref{BE}. The reason is presumably that
the {\em scattering} laws are the {\em same} in all models, and
equal to those of inelastic  {\it hard spheres} \Ref{coll}.

\noindent (v)  The eigenvalue $\lambda_s(\sigma)$ {\em does
depend} on the angular exponent $\sigma$.  We also recall that
$\lambda_s(\sigma)$ for the molecular chaos models has $\sigma
=1$. For the case $(d=3, \sigma=1)$ the eigenvalue reduces to a
simple expression,
\be
\lambda_s(\sigma=1,d=3) / \beta_1 = 1- \frac{2}{s+2} \left[
 (1-q)^s + \frac{1-q^{s+2}}{1-q^2}\right],
\ee
obtained before in Refs. \cite{BCG00,EB-EPL} for the
three-dimensional Maxwell model $(\nu=0, \sigma=1)$, satisfying
molecular chaos. To obtain this result form \Ref{38} one may use
the relation,
\be _2F_1(-\half s,1;2|z) = [z(1+\half s)]^{-1} \left[
 1 -(1-z)^{1+\half s}\right].
 \ee
One should also keep in mind that the $\nu$-dependence of
$\lambda_s(\sigma=\nu)$ in the simple model of Eq.\Ref{BE-nu} only
refers to the angular dependence of the collision rate
$K=|g_\parallel|^\nu=g^\nu |\hat{g}_\parallel|^\nu$.

\noindent (vi) The eigenvalue for $s=2$ has for all $d$ the simple
form
\be \Label{312}
\lambda_2(\sigma) = 2 p q \beta_{\sigma+2} \ ,
\ee
as follows from the relation $_2F_1(-1,b;c|z)=1-bz/c$ and
\Ref{A5}.

\noindent (vii) The behavior of $\lambda_s(\sigma)/\beta_\sigma$
as a function of $s$ is shown in Fig.\ref{fig2.1} for various
values of the inelasticity, where $\sigma=1$ and $d=2$.

Before concluding this section we analyze the asymptotic form of
the collision operator $I(c|f)$ at large $c$, in order to improve
the estimate, $I_\infty(c|f) \simeq -\beta_\sigma c^\nu f$, given
in the literature \cite{vNE-granmat}. This extension is needed to
obtain an asymptotic expansion of the form \Ref{51a}, or
equivalently,
\be \Label{315}
f(c)  \sim c^\chi \exp[-\beta c^b + \beta^\prime c^{b^\prime} ],
\ee
with $b>  b^\prime$. The analysis is given in \Ref{B7} of  Appendix
B, with the result for $c \gg 1$,
\be \Label{316}
-I_\infty (c|f)= \Lambda_\infty f(c) \simeq \beta_\sigma c^\nu \left[1
-{\cal K}_\sigma c^{-\half b(\sigma+1)}\right] f(c).
\ee
The coefficient is given by
\be \Label{317}
{\cal K}_\sigma     =\left\{ \begin{array}{ll}
\frac{\Gamma(\frac{d+\sigma}{2})}{\Gamma(\frac{d-1}{2})}
\left(\frac{2}{\beta b (1-q^2)} \right)^{(\sigma+1)/2}
& (\sigma \neq 1;\, \mbox{general})\\[3mm]
\frac{(d-1)}{\beta b(1-q^2)} & ( \sigma=1;\,
\mbox{molecular chaos})
\end{array} \right..
\ee
These relations are valid for all $\chi, \beta^\prime$, $b^\prime
< b $, and $d>1$. For $d=1$ there are only exponentially decaying
corrections, and ${\cal K}_\sigma =0$. The arguments, used here
and in the appendices, to analyze the properties of the operators
$\Lambda$ and $\Lambda_\infty$ are similar in spirit to those used
in Ref. \cite{EBN-Machta}.

\renewcommand{\theequation}{4.\arabic{equation}}
\setcounter{section}{3} \setcounter{equation}{0}

\section{Scaling equations}
\label{sec:IV}

As sketched  in a qualitative manner in Section II, the velocity
distribution function $F(v,t)$ shrinks --without external driving
and in zeroth order approximation-- to a delta distribution
$\delta(\bv)$ as the thermal velocity decreases. However, if one
re-scales the velocities in the shrinking distribution in terms of
its instantaneous width $v_0(t)$, then the rescaled distribution
rapidly approaches a {\it time-independent} scaling form $f(c)$,
\be \Label{41}
F(v,t) = (v_0(t))^{-d} f(c) \quad {\rm and} \quad c= v/v_0(t),
\ee
as discussed in Refs. \cite{EB-EPL,AB+CC-proof,EB-Springer}.
Direct Monte Carlo Simulations (DSMC) of the Boltzmann equation
\cite{Brey-scaling,MS00,Rome1,BBRTvW01} have confirmed that after
a short transient time  the rescaled velocity distribution can be
collapsed on a time-independent {\it scaling} solution $ f(c)$.
These observations indicate that the long time behavior of
$F(v,t)$ in freely cooling systems approaches a simple, and to
some extent {\it universal}, scaling form $f(c)$, which is the
same for a general class of initial distributions. It satisfies
the normalizations,
\be \Label{42}
\int d\bc \{1,c^2\} f(c) =\{1, \half d \},
\ee
consistent with the definition of $\int d\bv v^2 F(v,t) =\half d
v^2_0(t)$ of the r.m.s. velocity $v_0(t)$.

To study the scaling forms in the case of white noise driving, we
consider the equation for the mean square velocity, using the
Boltzmann equation \Ref{BE-driven} - \Ref{source},
\begin{eqnarray}
\Label{43}
\frac{d v_0^2}{dt} &=& \textstyle{\frac{2}{d}} \langle v^2|I(F)\rangle
+\textstyle{ \frac{2}{d}} D \langle v^2|\partial^2 F\rangle
\nn &=& -2
\gamma(\nu,\sigma) v_0^{\nu+2} +4 D.
\end{eqnarray}
Here, we have used \Ref{33} with $k=v^2$, and the relation $\Delta
E = -p q g_\parallel^2$ below \Ref{coll}, performed the rescaling
in \Ref{41}, and introduced
\be
\Label{44}
\gamma(\nu,\sigma) \,=\, \frac{1}{d}\, p q\, \beta_{\sigma+2} \,\langle\langle
|{\bf c}-{\bf c}_1|^{\nu+2}\rangle\rangle.
\ee
The average $\langle\langle...\rangle\rangle$ is calculated with
the time-independent weights, $f(c) f(c_1)$. So $\gamma
(\nu,\sigma)$ is an unknown constant, except for the cases $\nu=0$
(Maxwell) and $\nu=-2$ (WN-threshold model), where it takes the
values,
\be
\Label{45}
\gamma(0,\sigma) \,=\, p q \,\beta_{\sigma+2}
\qquad \hbox{and}
\qquad
\gamma(-2,\sigma) \,=\, p q \,\frac{1}{d}\,\,\beta_{\sigma+2}.
\ee
For the case of {\it free cooling} without energy supply ($D=0$),
the r.m.s velocity in \Ref{43} keeps decreasing for large $t$, as
\be \Label{46}
\frac{v_0(t)}{v_0(0)} \sim \left\{
\begin{array}{lr}
(1+\nu t/t_{c }(\nu,\sigma))^{-1/\nu}   & (\nu >0) \\
\exp[-t/t_{c}(0,\sigma)]               & (\nu = 0)    \\
(1-|\nu|t/t_{c}(\nu,\sigma))^{1/|\nu|}  &(\nu<0)
\end{array}
\right.,
\ee
where  the constant $t_c(\nu,\sigma) =1/[\gamma (\nu,\sigma)
v_0^{\nu}(0)]$ is unknown except for $\nu = \{0, -2\}$:
\be \Label{46a}
t_c(0,\sigma) \,=\, {1}/{pq \beta_{\sigma+2}} \qquad
\hbox{and}\qquad t_c(-2,\sigma) = {d}/{p q \beta_{\sigma+2}}.
\ee
Equation \Ref{46} is the extension of Haff's homogeneous cooling
law \cite{Haff} to inelastic soft spheres.

The most important time scale in kinetic theory is the inverse of
the mean collision rate, defined as the average of the collision
rate $K =g^\nu |\hat{g}|^\sigma$, i.e.
\begin{eqnarray}
\Label{410}
\omega(t) &=& \int_{\bf n} \int d\bv d{\bf w} \, g^\nu
|\hat g_\parallel|^\sigma \, F(v,t) F(w,t)
\nn &=& \beta_\sigma
\langle\langle |{\bf c}_1-{\bf c}|^\nu
\rangle\rangle \, v_0^\nu(t) \equiv \zeta(\nu,\sigma) v_0^\nu(t).
\end{eqnarray}
The second line above applies {\it only to scaling} solutions,
where $\zeta(\nu,\sigma)$ is known explicitly for $\zeta(2,\sigma)
= d \beta_\sigma$  (Very Hard Particles) and $\zeta(0,\sigma)=
\beta_\sigma$ (Maxwell). The collision frequency $\omega(t)$
depends on the external or laboratory time $t$. We also consider
the collision counter or internal time of a particle $\tau$,
defined through the relation $d
\tau = \omega(t) dt$. It represents the total number of collisions
$\tau$ that a particle has suffered in the external time $t$. With
the help of \Ref{410} and \Ref{46} this differential equation can
be solved to yield
\be \Label{411}
\tau = \frac{1}{\varpi(\nu,\sigma)} \ln \left[ 1 + \nu
\frac{t}{t_c(\nu,\sigma)} \right]^{1/\nu},
\ee
where
\be \Label{412}
\varpi(\nu,\sigma) \,=\, \frac{\gamma(\nu,\sigma)}{\zeta(\nu,\sigma)}
\,=\, \frac{p q \,\beta_{\sigma+2}\, \langle\langle |{\bf c}-{\bf c}_1|^{\nu+2}
\rangle\rangle}{d \,\beta_\sigma\,\langle\langle |{\bf c}-{\bf c}_1|^{\nu}
\rangle\rangle},
\ee
is an unknown constant, as it depends on $f(c)$. It reduces for
$\nu=0$ to $\varpi(0,\sigma) = p q
\beta_{\sigma+2}/\beta_\sigma$. By combining \Ref{411} and
\Ref{46} in the case of free cooling, the homogeneous cooling law
takes the simple exponential form
\be
\Label{413}
v_0(t) \,=\, v_0(0) \, \exp[-\varpi(\nu,\sigma)\tau].
\ee
On the other hand for $\nu<0$ the r.m.s. velocity $v_0(t)$ in
\Ref{46}, and the granular temperature $v_0^2(t)$ vanish in a
finite time $t_c(\nu,\sigma)/|\nu| = 1/\gamma (\nu,\sigma)
v_0^{\nu}(0)$. At the same time the collision counter $\tau $ in
\Ref{411} diverges. This means that in inelastic soft sphere
models with $\nu <0$  an {\it infinite} number of collisions
occurs in a {\it finite} time $t_{c}(\nu,\sigma) /|\nu|$. A
collision counter, diverging within a finite time, is also
observed  in the phenomenon of inelastic collapse \cite{McNam+Y},
occurring in inelastic hard sphere fluids. There a finite number
of particles line up in a linear array in configuration space. The
divergence of the collision counter $\tau$ in \Ref{411} at
$t=t_c(\nu,\sigma)/ |\nu|$ is a reflection in velocity space of
this {\it inelastic collapse phenomenon}. In the WN-driven case
one may also analyze the collision counter $\tau$ through $d\tau =
\omega(t) dt$, and one observes similar phenomena of inelastic
collapse for $\nu<-2$.

Next we consider the NESS. In the case of WN-driving the energy
balance in \Ref{43} has a stationary or fixed point solution
$v_0(\infty)$, given by
\be\Label{47}
v_0^{2b}(\infty) \,=\, {2D}/{\gamma(\nu,\sigma)} \qquad (2b =
\nu+2),
\ee
and the energy balance equation \Ref{43} can be expressed as,
\be
\Label{48}
\frac{d v_0^2(t)}{dt} \,=\, 4 D \left[
1-\left(\frac{v_0(t)}{v_0(\infty)}\right)^{2b}
\right].
\ee
The solution $v_0(t)$ may approach to or move away from
$v_0(\infty)$, depending on the sign of $b$. For $b=1+\nu/2>0$,
the fixed point solution is stable and attracting, for $b<0$ it is
unstable and repelling, and for $b=0$ it is marginally stable. In
a system at the stability threshold ($b=0$), a  marginally stable
NESS only exists if one can fine tune the driving parameters in
\Ref{43} to $D=\gamma(-2,\sigma)/2=p q \beta_{\sigma+2}/d$.
Otherwise, the energy for $b=0$ or $\nu=-2$ is increasing or
decreasing linearly with time,
\be \Label{49}
v_0^2(t) = v_0^2(0) + (4D-2pq \beta_{\sigma+2}/d)t.
\ee
The possible existence of time-independent scaling solutions can
be investigated by substituting \Ref{41} into \Ref{BE} with the
result
\begin{eqnarray}
\Label{414}
I(c|f) &=& -\frac{\dot v_0}{v_0^{\nu+1}} \,\partial \cdot({\bf
c}f) -\frac{D}{v_0^{\nu+2}} \,\partial^2 f \nn
&=&\left\{\gamma(\nu,\sigma) -\frac{2D}{v_0^{\nu+2}}\right\}
\partial\cdot({\bf c}f) -\frac{D}{v_0^{\nu+2}} \,\partial^2 f.
\end{eqnarray}
On the last line $\dot v_0$ has been eliminated using \Ref{43}.
This equation is not yet an integral equation for a scaling
function, because the coefficients still depend on time through
$v_0(t)$.

There are two possibilities for time-independent solutions. The
first one is obtained by setting $D=0$. This gives the integral
equation for the scaling solution in the free cooling state, which
has been studied extensively in the literature
\cite{vNE-granmat,EB-Rap,EB-Springer}, and leads to high energy
tails of stretched exponential form for $\nu>0$ or to power law
tails for $\nu=0$. The second possibility for a time-independent
solution is for the stable NESS in the WN-noise driven case with
$v_0(\infty)$ given by \Ref{47}. Then the coefficient of the first
term on the r.h.s. of \Ref{414} vanishes, and the integral
equation reduces to,
\be
\Label{415}
I(c|f) = - \half \,\gamma(\nu,\sigma) \,\partial^2 f \,= - \half
\,\gamma(\nu,\sigma)\left(f''+\frac{d-1}{c} f'\right)
\ee
where a prime denotes a $c-$derivative.

However, at the stability threshold ($\nu =-2$) there exists a
marginally stable state, described by \Ref{49} with $v_0(\infty) =
v_0(0)$, and obtained by fine tuning the driving parameter to the
value $D =\half \gamma(-2,\sigma)$. This yields the scaling
equation for the WN-noise threshold model with $\nu =-2$,
\be
\Label{416}
I(c|f) = - \frac{1}{2d} p q\,\beta_{\sigma+2} \partial^2 f =
-\frac{1}{4d} \lambda_2(\sigma) \partial^2 f ,
\ee
where \Ref{45} and \Ref{312} have been used. Furthermore, it is
interesting to note that the existence of a time independent
scaling form does {\it not} require the energy $v_0^2$ to be
stationary in the WN-threshold model ($\nu=-2$). Suppose we add an
additional friction force ${\bf a} =\gamma_0 \bv$, the Gaussian
thermostat, to the source term of the Boltzmann equation
\Ref{BE-driven}, (where $\gamma_0$ may be positive or negative),
then the energy balance equation \Ref{43} and the integral
equation for the scaling form \Ref{414} both get an extra term,
i.e.
\ba \Label{417}
v_0 \dot{v}_0 &=& - \gamma (-2,\sigma) +2D +\gamma_0 v_0^2 \nn
I(c|f) &=& [-v_0 \dot{v}_0 +\gamma_0 v^2_0\,]  \partial \cdot (\bc
f) -D \partial^2 f
\nn &=& [\gamma (-2,\sigma) - 2D \,] \partial \cdot (\bc f)
 -D \partial^2 f.
\ea
When $\dot{v}_0$ is eliminated from the first two equations, both
extra terms containing $\gamma_0 v_0^2 $ cancel, and the resulting
equation is identical to \Ref{414} with $\nu =-2$. This is an
integral equation for the scaling form.

Subsequent fine tuning of the diffusion coefficient to the value
$D = \half \gamma (-2,\sigma)$ simplifies the integral equation to
the same scaling equation \Ref{416}, and the energy balance
equation to $\dot{v}_0 = \gamma_0 v_0$. Consequently in the limit
of large time energy may  diverge or vanish, rather than stay
constant. So, the {\it total} energy does {\it not} have to be
{\it finite} for the existence of a scaling form, and the time
independent scaling solution $f(c)$ of the fine-tuned WN-driven
Boltzmann equation in \Ref{416} is not affected by adding an extra
Gaussian thermostat.

In performing DSMC simulations it should be noted that the
realization of a marginally stable NESS  is more complicated for
driving by a {\it stochastic force} than by a {\it deterministic}
friction force. The reason is that $D$ is the mean strength of the
random kicks $\av{\xi^2}$, and the realized strength  fluctuates
around the mean. So fine tuning the value of $D$ by choosing it to
satisfy \Ref{47}, does not make $\dot{v}_0$ vanish exactly, but it
may be a bit positive or negative. Then by re-scaling the
velocities respectively down or up at regular intervals, i.e. by
using the Gaussian thermostat, ${\bf a} = -\gamma_0 \bv$,
respectively with $\gamma_0 >0 $ one can decrease $\dot{v}_0$ to
0, or with $ \gamma_0<0$  increase $\dot{v}_0$ to 0, thus reaching
the NESS. This addition of an extra Gaussian thermostat leaves the
integral equation for the scaling function invariant, as we have
seen in \Ref{417}. One may also consider an
alternative but very similar algorithm, as given in Ref.\cite{Gamba3}. 

Exact closed form solutions of the scaling equations are not
known, except for the freely cooling one-dimensional Maxwell model
\cite{Rome1}. However, the high energy tails of $f(c)$ can be
calculated explicitly with the new methods, presented in this
article, both for exponentially bounded tails, $\exp[-\beta c^b]$
appearing in stable NESS, as well as for power law tails, $f(c)
\sim c^{-a}$ appearing in marginally stable NESS ($b=0$).

\renewcommand{\theequation}{5.\arabic{equation}}
\setcounter{section}{4} \setcounter{equation}{0}
\section{Asymptotics for stable NESS ($\nu>-2$) }
\label{sec:V}

In this section we focus on the high energy behavior of the
scaling form $f(c)$ for a stable NESS ($b=1+\nu/2>0$) for
inelastic soft sphere models as determined by the integral
equation \Ref{415}. For large $c$-values the collision operator
reduces to $I(c|f) \sim -\beta_\sigma c^\nu f(c)$ and the solution
of \Ref{415} has the form $f(c) \sim \exp(-\beta c^b)$ with
$b=1+\nu/2$, as has been derived in Refs
\cite{vNE-granmat,EB-Rap,EB-Springer}. The goal of this section is
to determine the sub-leading correction in the asymptotic
expansion \Ref{51} of $\ln f(c)$ for large $c$, i.e.
\be \Label{51}
\ln f(c)\simeq  -\beta c^b \,+\beta' c^{b'} +\chi \log c + \log A
+ ...
\ee
where $b>b'$. To determine these exponents and coefficients we need
a better asymptotic estimate of the Boltzmann collision operator
$I(c|f)$, when acting on a function of the form \Ref{315}.

According to Appendix B the large$-c$ behavior of $I(c|f)$ is
determined by the {\it linear } operator $\Lambda_\infty$, which is
given in \Ref{B1}. It is a limiting form for $ c \gg1$ of the
linearized Boltzmann collision operator $\Lambda$, and its action
on functions of the form \Ref{315} has been analyzed in Appendix B
with the result \Ref{B7},
\be \Label{52}
I_\infty(c|f) \,=\, -\Lambda_\infty \,f \, \simeq \,
\beta_\sigma \,c^\nu \,\left[1-{\cal K}_\sigma \,c^{-a} \right]
 \,f(c),
 \ee
where ${\cal K}_\sigma$ is a constant given in \Ref{B6} and $a =
 b(\sigma+1)/2$. Then the scaling equation \Ref{415} becomes
 \be
\Label{53} \beta_\sigma \,c^\nu \left[1-{\cal K}_\sigma \,c^{-a}
\right] f\,=\, \half \gamma(\nu,\sigma) \left\{ f''+
\left(\frac{d-1}{c}\right)\,f' \right\}.
\ee
As this equation is homogeneous, the constant $A$ in \Ref{51} cannot be
determined. So we set $A=1$ and for consistency impose the
restriction $b>b'>0$, as $c^{-|b'|} \ll |\ln  A|$ for $c \gg 1$.
 The parameters in \Ref{51} can be determined
by substituting $f(c)$ of \Ref{315} into \Ref{52} and matching the
exponents of powers of $c$ and equate the coefficients, i.e.
\be
\Label{54}
\eta_\sigma c^\nu \left[1-{\cal K}_\sigma \,c^{-a}
\right] \,\simeq \, \beta^2 b^2\, c^{2b-2} \,-\, 2 \beta b \,
\beta' b'\,c^{b+b'-2} \,-\, \beta b\,c^{b-2}
\left[d-2+b+2\chi \right] \,+\, ...
\ee
To leading order we equate the terms proportional to $c^\nu$ and
$c^{2b-2}$, which yields the exponent $b$ and the coefficient
$\beta$, with  the well known results \cite{EB-Springer},
\be \Label{55}
f(c) \sim \exp[-\beta c^b]\quad, \quad b=1+\half{\nu}>0 \quad,
\quad b\beta  = {\eta_\sigma}\equiv \sqrt{ 2 \beta_\sigma /\gamma
(\nu,\sigma)}.
\ee
The dominant tails are stretched Gaussians ($0<b<2$), or
compressed ones ($b>2$). These tails are respectively
over-populated or under-populated when compared to a Gaussian. The
results for soft sphere models with $b=1+\half\nu>0$ are valid for
all dimensions $d \geq 1$. We also note that the exponents $b$
found here are equal to the $b$-values that determines the
stability threshold in the energy balance equation \Ref{48}. In
principle, $\gamma(\nu,\sigma)$ in \Ref{55} can be calculated
perturbatively, using the method developed in Ref
\cite{vNE-granmat,EB-Springer}, or it can be measured
independently using DSMC, as will be done in the present paper.

We first consider Eq. \Ref{54} for the molecular chaos models with
$\sigma=1$, where \Ref{B6} gives $\beta b {\cal K}_1 =
({d-1})/({1-q^2})$, and $a=b=1+\half\nu$. The only consistent matching
of the sub-leading exponents (satisfying the restriction $b'>0$) is
obtained by setting $\beta'=0$. The parameters in the sub-leading
correction are then
\be \Label{56}
\beta'=0 \quad, \quad \chi =- \frac{1}{2} (d-1) - \frac{1}{4} \nu + \frac{1}{2}
\beta b{\cal K}_1   = \frac{(d-1) q^2}{2(1-q^2)} -
\frac{1}{4} \nu
\ee
valid for all allowed $q$-values $0<q=\half(1-\alpha) < \half$.
For $\sigma\neq 1$ one can easily verify that the exponents in
\Ref{54} obey the relation $\nu-a \neq b-2$, and matching gives
either $\nu-b=b+b'-2>b-2$ or the reversed inequality. In the
former case, $b'=b(1-\sigma)/2>0$ which is realized for
$\sigma<1$. Then equating coefficients yields $\beta'=\beta{\cal
K}_\sigma/(1-\sigma)$. In this case the sub-leading asymptotic
correction in \Ref{51} is $\beta' c^{b'} \gg |\chi| \,\log c$, and
in the spirit of asymptotic expansions we set $\chi=0$.

For the reversed inequality $b'=b(1-\sigma)/2<0$, realized for
$\sigma>1$, the sub-leading asymptotic correction in \Ref{51} is
$|\chi| \log c \gg \beta' c^{b'}$, and in the same spirit as
above, we set $\beta'=0$. The value of exponent $\chi$ is found by
setting the coefficient of $c^{b-2}$ in \Ref{54} equal to zero. In
summary:
\ba \Label{58}
\beta'=\,0,\qquad &\chi = - \half(d-1)- \textstyle{ \frac{1}{4}}
\nu &\qquad (\sigma>1)\\
\Label{57}
b' = \half b(1-\sigma),\qquad &\beta' = \beta\, {\cal
K}_\sigma/{(1-\sigma)}, \qquad \chi=0,& \qquad (|\sigma|<1)
\ea
Next we focus  for the WN-driven case on comparison of the Monte
Carlo simulations with the theoretical predictions.
Fig.\ref{fig3.4} shows the DSMC data of one-dimensional soft
sphere models for $f(c)$ as a function of $c^b$ with $b=1 +\half
\nu$ for several values of $\nu$. Moreover, the asymptotic
large$-c$ prediction, $\ln f(c) \sim -\beta c^b -(\nu/4) \ln c$ in
\Ref{56}, is confirmed in Fig.\ref{fig3.5}. The value of $\beta$
is independently obtained by measuring $\gamma(\nu,\sigma)$ in
\Ref{55} during the DSMC simulation. Note that for the
one-dimensional Maxwell model the prediction $f(c) \sim \exp[- c
\sqrt{2/pq} ]$ has a very simple form, because \Ref{55} gives
$\eta_0 =\sqrt{2/pq}$ as $\beta_s =1$ for all $s$. Moreover, this
figure shows that the  correction, $-(\nu/4) \ln c$, leads to an
improved agreement between theory and simulations, even though it
is tiny because of the limited range of velocities accessible.

The two-dimensional WN-driven model, with $\sigma=\nu=- 1/2$, is
illustrated in the next two figures. Fig.\ref{fig3.6} first
displays the measured scaling form $f(c)$, plotted versus $c^b$
with $b=1+ \nu/2 =3/4$ for different values of $\alpha$. The
straight parts of the high energy tails confirm the dominant
velocity dependence, $\ln f(c) \sim -\beta c^{3/4}$, with $\beta$
increasing with $\alpha$. For large enough $\alpha$, $f(c)$ is
getting closer to a Gaussian. Moreover, Fig.\ref{fig3.6ab}
presents a very interesting illustration of the sub-leading
correction $\exp(\beta' c^{b'})$. While a plot of $\ln f(c)$
versus $c^b$ yields a straight line, and thus an apparent
agreement with the dominant theoretical prediction $\exp(-\beta
c^b)$, closer examination shows that measuring $\beta$ through a
fit to $\exp(-\beta c^b)$ would disagree with the predicted value
of $\beta$ in \Ref{55}, which is computed from the numerical data
using the definition (\ref{55}) which involves
$\gamma(\nu,\sigma)$. On the other hand, comparison of $f(c)$ with
the full expression $\exp(-\beta c^b+\beta' c^{b'})$ --where
$\beta'$ is measured from its definition (\ref{57}) involving
${\cal K}_\sigma$-- gives a very good agreement.

Regarding the soft sphere systems in stable non-equilibrium steady
states with model parameters $\nu > -2$, we may conclude that the
agreement between analytic and DSMC results for high energy tails
is very good. Essentially all analytic and numerical results  in
this section are {\it new}. They are the leading and sub-leading
asymptotic correction factors in the scaling function $f(c)$   in
a stable NESS ($b>0$) for the general class of inelastic soft
sphere models. These results include the few special cases known
in the literature, i.e. the dominant asymptotic form $f(c)\sim
\exp[-\beta c^b]$ for white noise driven inelastic hard spheres
$(\nu=1)$ and similar results for inelastic Maxwell models
$(\nu=0)$. The only known result about sub-leading correction
factors concerns the very special case of the three-dimensional
molecular chaos Maxwell model $(d=3,\nu=0,\sigma=1)$, for which
Bobylev and Cercignani \cite{BC02-JSP}   have derived the result
$f(c) \sim c^\chi \exp[-\beta c^b]$ with $\chi = q^2/(1-q^2)$.
Although the derivation is not very transparent, their result
agrees with \Ref{56} for this special case.

\renewcommand{\theequation}{6.\arabic{equation}}
\setcounter{section}{5} \setcounter{equation}{0}
\section{Power law tails at marginal stability ($\nu=-2$)}
\label{sec:VI}
Marginal stability is a limiting property of a stable NESS as $b
\to 0^+$, which can not occur in free cooling, but only in {\it
driven} states, here driven by white noise.  Below \Ref{49} we have
explained that the marginally stable NESS can not be reached by
natural time evolution, but only by fine tuning the external
parameters in the energy source term.

As we have seen in \Ref{55}  the asymptotic solution $f(c)$ for
stable states ($b=1+\half \nu > 0$) is to leading order described
by $f(c) \sim A \exp[-\beta c^b]$ with $\beta={\eta_\sigma}/b $.
To illustrate how power law tails arise, it is convenient to set
$A=e^\beta$, and take the limit of $f(c)$ as $b \to 0$. The result
is $f(c) \sim \exp [-\eta_\sigma (c^b-1)/b] \to
c^{-\bar{\eta}_\sigma}$, with $\overline{\eta}_\sigma =\lim_{b \to
0}\eta_\sigma$. Of course $\overline{\eta}_\sigma$ is not the full
exponent of the tail, because the exponential form above
represents only the leading asymptotic behavior for $b>0$. For
instance, any correction factor $\exp[-\beta' c^{b'}]$ as
appearing in \Ref{315}, where $b'=B(b) \to 0$ as $b \to 0$, would
give additional contributions.

To determine the exact exponent of the power law tail we have to
solve the scaling equation \Ref{416} for the WN-threshold model,
where $\nu =-2$. As we expect power law solutions from the
arguments above, we use the linearization in \Ref{29}-\Ref{31} to
give,
\be \Label{61}
-I(c|\delta+h) \simeq \Lambda h = \frac{1}{4d} \lambda_2(\sigma)
\partial^2 h = \frac{1}{2d} p q\beta_{\sigma+2}
\partial^2 h,
\ee
where $f(c) \sim h(c)$ for $c\geq 1$. Inspection of this equation
shows that the operators on both sides of the equation  have the
same set of R-eigenfunctions, $c^{-s-d+2}$. Substituting this
function into \Ref{61} gives the {\it transcendental} equation,
\be \Label{62}
\lambda_s(\sigma)= (1/4d) s(s+d -2) \lambda_2(\sigma)= (1/2d) s(s+d-2)
pq \beta_{\sigma+2}.
\ee
The possible roots of this equation, $s=a_d(\alpha)$, depend on
the spatial dimensionality $d$, and on the coefficient of
restitution $\alpha$ through $p=1-q =\half(1-\alpha)$. They are
candidates for power law exponents in an asymptotic solution of
the form,
\be \Label{63}
f(c) \sim c^{-a-d + 2}
\ee
for the WN-threshold model. However there exist a priori
restrictions on the exponent $a$ through the normalizations
\Ref{42}. If one is interested in solutions with bounded energy,
then the constraint,  $\int d\bc c^2 f(c) < \infty$,  or
equivalently $a>4$, applies for all $d$. If one allows the energy
to keep increasing -- as discussed below \Ref{417}-- one still has
to obey the restriction, $\int d\bc f(c) <\infty$, implying $a>2$
for all $d$. Moreover, for $d>1$ the restriction $ \sigma > -1 $
applies, as discussed below \Ref{BE-nu}. At $d=1$ the angular
exponent $\sigma $ is absent.

To start we consider the one-dimensional case, where Eq.\Ref{62}
reduces to $1-p^s -q^s =\half s(s-1) pq$. It has two solutions: $
s_L=1$ and $s_R=a=3$. Only the largest root satisfies the
normalization constraint, and corresponds to the solution, $f(c)
\sim 1/c^{a+d-2} \sim 1/c^2$. We also note that in the
one-dimensional case the power law exponent $a$ is {\it
independent} of the coefficient of restitution. In our DSMC
calculations the energy is kept bounded by applying the Gaussian
thermostat, which does not affect the integral equation for the
scaling function, as discussed at the end of Section IV.

This  analytic result is in good agreement with the
one-dimensional DSMC results as shown in Fig.\ref{fig4.4}, where
the algebraic tail is observed over more than 3 decades in $f(c)$. The
energy in this state is however infinite. Why this high
overpopulation of the power law tail? The WN-threshold model is
very inefficient in equilibrating its {\it high energy} particles
in comparison with hard spheres ($\nu=1$), and even with Maxwell
models ($\nu=0$), because the tail particles rarely suffer a
collision as their collision rate decreases as $  K \sim |c|^\nu
\sim 1/|c|^2$.

To obtain a qualitative understanding of the $\alpha-$ and
$d-$dependence of the exponent $a_d(\alpha)$, we use a graphical
solution method by plotting the l.h.s. $y_s$ and the r.h.s.
$\bar{y}_s$ in \Ref{62} as functions of $s$, and determine the two
intersection points. Here we have defined,
\be \Label{64}
y_s = \frac{\lambda_s(\sigma)}{\beta_\sigma} \quad \mbox{and} \quad
\bar{y}_s = \frac{pq( \sigma+1)}{2d(\sigma +d)   } s(s+d-2),
\ee
where the relation $\beta_{\sigma +2}/\beta_\sigma = (\sigma+1)/(
\sigma+d) $ has been used. The procedure is illustrated in
Fig.\ref{fig4.5} for $(d=2,\sigma =1)$. In the elastic limit
$(\alpha \to 1^-)$ the pre-factor in $\bar{y}_s$ becomes smaller,
and the right intersection point, $s_R =a_d(\alpha)$, moves to
infinity. It satisfies the condition $a>4$, and is an acceptable
power law exponent. The left intersection point, $s_L$, remains
less than 2, and does not satisfy the constraints of finite energy
and mass.

The {\it numerical} solution of \Ref{62} for $(d=2, \sigma=1)$
yields  $a_d(\alpha)$,  shown in Fig.\ref{fig4.6}.
Fig.\ref{fig4.7} shows the good agreement between the predicted
power law tail and the DSMC measurements of $f(c)$ for $d=2$ and
various values of $\alpha$.

Guided by the numerical result, $a_2(\alpha=0) \simeq 4$ , we have
verified that this value is an exact solution of \Ref{62} for
$d=2$.  To do so we have calculated $\lambda_4(1)$ using the
relation $_2F_1(-2,1,\half(d+1)|3/4)=(d^2+d-3/2)/(d+1)(d+2)$, as
can be obtained from the second degree polynomial
$_2F_1(-2,b;c|z)$ in $z$, given through the series in \Ref{A8}
that terminates after the term with $n=2$. For higher dimensions
no such simple solutions seem to exist. For instance,
$a_d(\alpha=0) = 4.929, 5.812, 6.672 ...$ for $d=3,4,5,...$. As
$a_d(\alpha)$ is an increasing function of $\alpha$, our graphical
and numerical analysis shows that the power law tails, found in
the present section for $ d\geq 2$, all carry a finite energy.

For general $d$ the transcendental equation \Ref{62} can only be
solved in a few limiting cases. By taking the {\it elastic} limit
($q \to 0$) in \Ref{62}, and comparing right and left hand sides
one sees that the solution $s$ must become large as $q \to 0$. So
we need $\lambda_s(\sigma=1)$ for large $s$, as given in
\Ref{311}, and the dominant behavior of $a$ at small $q$ is,
\be \Label{65}
a_d(\alpha) = \sqrt{2d(d+\sigma)/q(1+\sigma)} \qquad (q \to 0; \,
d>1).
\ee
further sub-leading corrections to \Ref{65} can also be calculated
using asymptotic results for $_2F_1(a,b;c|z)$ at large $a$.

For {\it large dimensionality}  we take the coupled limit as $d
\to \infty$ and $s \to \infty$ while keeping $x= s/d =$ fixed, and
we set $\sigma=1$. This leads to an algebraic equation of third
order with three roots $\{x_- <0, x_0=0, x_+>0 \}$, where the
largest one determines the exponent in the power law tail, and
reads,
\be \Label{66}
a_d(\alpha)= x_+ d \sim \frac{d}{2q(1-q^2)}
\left\{\sqrt{4q(1+q)(1-q^2) +q^6 } -2q +q^3 \right\}  \qquad (d
\to \infty).
\ee
The behavior of $x_+$ vs $\alpha$ is illustrated in
Fig.\ref{fig4.6}. The results \Ref{65} and \Ref{66} agree to
dominant order in the respective limits $d \to \infty$ and $q \to
0$.

Finally, we  discuss the power law tail generated by an energy
source ``at infinity'', as recently proposed in Ref.
\cite{EBN-Machta}. Here the heating device injects energy in the
ultra high energy tail of the v.d.f. It works as follows. By
randomly selecting a particle at a very small rate $\gamma \ll
\omega_0$ (being the mean collision rate), and giving the selected
particle a (macroscopic) amount of energy $\Delta E(\gamma)$,
equal to the total energy lost by the system in the typical (long)
interval $1/\gamma$ between two injections. After a short
transient time, the energy cascades down to lower and lower
energy, and in that manner a NESS-v.d.f $f(c)$ is maintained that
does not evolve under inelastic collision dynamics.

The mathematical framework developed in the present paper, is very
suitable to discuss the problem posed above. We are in fact
looking for a solution of equation $I(c\,|f)=0$. This implies
through Eqs. \Ref{30}-\Ref{31} that we want a solution of $\Lambda
h(c)=0$. The answer is given by \Ref{314}. The required scaling
function is the $R$-eigenfunction $h(c)\sim c^{-s-d-\nu}$, where
$s$ is the root of $\lambda_s(\sigma)=0$ with $\lambda_s$ given in
\Ref{39}. This expression is identical to Eq. (11) of Ref.
\cite{EBN-Machta}, as can be verified from \Ref{A15}.

We recall from property (iii) below \Ref{311} that
$\lambda_s(\sigma)$ is independent of $\nu$-exponent in the
collision rate $K\sim g^\nu |\cos\theta|^\sigma$, and the
$\sigma$-exponent refers only to the angular dependence. For
molecular chaos models $\sigma=1$, as shown in \Ref{P-b}, and we
restrict ourselves to this case.

The graphical solution to this problem can be seen in
Fig.\ref{fig2.1}, as the point of intersection of the
$\lambda_s$-curve and the $s$-axis. For $d=1$, the eigenvalue
reduces according to \Ref{37} to $1-p^s-q^s=0$ with solutions
$s=a^*=1$, and the scaling form $f(c)\sim c^{-a-d-\nu}\sim
c^{-2-\nu}$. For dimensions $d\geq 2$ the scaling form has the
power law tail,
\be
f(c) \sim \frac{1}{c^{a^*+d+\nu}} \label{67}
\ee
where $s=a^* = a_d^*(\alpha)$ is the solution of
$\lambda_s(\sigma=1) = 0$ and depends on $d$ and $\alpha$.

From general considerations one obtains the bounds
\cite{EBN-Machta}
\be
1 \,=\, a^*_1(\alpha) \,\leq \, a^*_d(\alpha) \,\leq \, a_d^*(1)=2,
\ee
where the upper bound is related to energy conservation. The
numerical solution of $\lambda_s=0$ yields the curves in
Fig.\ref{fig5}. The variations in $a_d^*(\alpha)$ with $\alpha$
and $d$ are small. It is interesting to note that not all
solutions have finite energy: indeed the constraint
$\int d{\bf c}\, c^2 f(c) <\infty$
implies for all $d$ and all $\alpha$ the bound
$a_d^*(\alpha)> 2-\nu$. Consequently, for inelastic hard spheres
($\nu=1$) the stationary solution with the tail \Ref{67} has a
finite energy for all $d\geq 2$, whereas  in the softer Maxwell
models ($\nu=0$) the solutions all carry infinite energy.

Analytic results for the present model are very limited. Near the
elastic limit ($q$ small), one finds the exponent $a^* = 2-x$,
where $x\simeq {\cal O}(q)$,  by expanding the terms in
$\lambda_s$ for small $x$. The result after some calculations
reads
\be
a^*_d(\alpha) \,=\, \frac{4q}{\gamma_E +\psi(\frac{d+1}{2})} \,+\,
{\cal O}(q^2),
\ee
where $\gamma_E \simeq 0.5771...$ is Euler's constant and
$\psi(z)=\Gamma'(z)/\Gamma(z)$ is the logarithmic derivative of
the Euler Gamma function, the so-called Digamma function.

\renewcommand{\theequation}{7.\arabic{equation}}
\setcounter{section}{6} \setcounter{equation}{0}
\section{Conclusions and perspectives}
\label{sec:VII}

In the present paper we have studied asymptotic properties of
scaling or similarity solutions, $ F(v,t) = (v_0(t))^{-d}
f(v/v_0(t))$, of the nonlinear Boltzmann equation in spatially
uniform systems composed of particles with inelastic interactions
for large times and large velocities. The large $t-$ and $c-$
scales are relevant because on such scales the universal features
of the solutions survive, as the velocities, $c= v/v_0(t)$, are
measured in units of the r.m.s. velocity or instantaneous width
$v_0(t)$ of the distribution. It follows from DSMC solutions that
the v.d.f. $F(v,t)$ for large classes of initial distributions,
possibly driven by energy sources, 
 evolve into a scaling solution, and  after a
sufficiently long time the combination $v_0^d(t) F(v,t) $, plotted
versus $c = v/v_0(t)$,  can be collapsed for all $t$ on a single
scaling form $f(c)$  with an overpopulated high
energy tail. This observation has been rigorously
proven for Maxwell models in Ref. \cite{AB+CC-proof},
and for hard spheres in Ref.\cite{Gamba1}.

In this article we have focused on the properties of scaling
solutions, and in particular on its high energy tail. In Section
\ref{sec:II} we have introduced the Boltzmann equation for classes
of inelastic interactions, corresponding to pseudo-repulsive power
law potentials, $V(r) \sim 1/r^n$, with collision rates scaling
like $K \sim g^\nu$ where $g$ is the relative speed of the
interacting particles.  These models embed hard scatterers like
inelastic hard spheres ($\nu =1$) and soft scatterers like
pseudo-Maxwell molecules ($\nu=0$), and even softer ones with $\nu
<0$. The energy loss in an inelastic interaction is proportional
to the inelasticity, $(1-\alpha^2)$, where $\alpha$ is the
coefficient of restitution. We also show in Section \ref{sec:II}
how driving forces, such as white noise or nonlinear friction can
be included in the Boltzmann equation.

Section \ref{sec:III} and Appendices \ref{app:A} and \ref{app:B}
provide the crux of the new method, that enables us to determine
the {\em singular} large$-c$ part $h(c)$ of the scaling form
$f(c)=\delta({\bf c}) + h(c)$. The basic observation is that
$f(c)=\delta({\bf c})$ is invariant under collisions, i.e.
$I(c|\delta)=0$. The subsequent linearization in Eqs.
\Ref{30}-\Ref{31} of the nonlinear Boltzmann equation around this
singular stationary solution provides the linearized collision
operator $\Lambda$ that determines $h(c)$.

In the Fourier transformation method --which can only be applied
to Maxwell models-- the corresponding Fourier transforms are
$\widehat f(k) \,=\, 1+\widehat h(k)$ and $\widehat I(k|\widehat
f)$, where ${\bf k}$ is the variable conjugate to ${\bf c}$. Then,
one has $\widehat I(k|1)=0$, and it leads to the linearized
operator \smash{$\widehat I(k|1+\widehat h) \simeq
-\Lambda^\dagger \widehat h(k)$}, where $\widehat h(k)$ has a
leading small $k$ singularity of ${\cal O}(k^s)$, where $s$
differs from an  even positive integer. This method leads directly
to the eigenvalue equations, solved in Refs.
\cite{BNK-JPA02,EB-EPL} to determine the power law exponent $a$ in
$f(c) \sim 1/c^{a+d}$ for Maxwell models. The arguments above, in
the reverse direction, have generated the essential suggestion to
introduce $\Lambda$ for general soft sphere models with $\nu\neq
0$, and to study its properties.

Section \ref{sec:IV} provides the important link between on the
one hand the stability ($b>0$) and the marginal stability ($b=0$)
of fixed point solutions of the energy balance equation,  and on
the other hand the approach of solutions of the Boltzmann equation
to a scaling form $f(c)\sim \exp(-\beta c^b)$ with the stretched
exponent $b>0$,  as well as the existence of power law solutions
$f(c) \sim c^{-a-d-\nu}$ in the $\nu$-model at its stability
threshold $(b=0)$.

The stretching exponent for the $\nu$-models with white noise
driving, discussed in this paper, is $b=1+\nu/2$. The $b$-values
for free cooling and general nonlinear friction were mentioned in
the introduction.   The existence of power law solutions in the
corresponding threshold models $(b=0)$, as well as the calculation
of the power law exponents will be published elsewhere
\cite{ETB-II}. The application of the general theory to white
noise driving for stable soft sphere models was presented in
section \ref{sec:V} where also the importance of sub-leading
corrections was demonstrated in Fig.\ref{fig3.6ab}.

We have also analyzed an inelastic $d$-dimensional BGK model
\cite{Brey-BGK,EB-Springer} for free cooling and white noise
driving. In the former case the scaling solution can be solved
exactly. The model shows an algebraic tail $f(c)\sim c^{-a-d}$
with an exponent $a\sim 1/(1-\alpha^2)$, where $1-\alpha^2$ is
again the fractional loss of energy in a collision. For white
noise driving, one finds asymptotically $f(c)\sim \exp(-\beta c)$.
So the BGK model exhibits the generic behavior of a Maxwell model.

In section \ref{sec:VI} the power law exponent $a=a_d(\alpha)$ in
the threshold model $(b=1+\nu/2=0)$ for white noise driving is
determined from a transcendental equation that can be solved
exactly for $d=1$ and leads to  the power law tail $f(c)\sim
c^{-a-d-\nu}\sim c^{-2}$, as illustrated in Fig.\ref{fig4.4}. For
general dimensions, the transcendental equation has been studied
analytically, graphically and numerically and the power law
exponent $a_d(\alpha)$ for $d=2$ and $\sigma=1$ is compared with
the asymptotic result for large $d$ in Fig.\ref{fig4.6}. In our
new asymptotic method the power law tail for a system driven by
the ultra-high energy source is simply determined by the
eigenvalue equation, $\Lambda h_s(c) =\lambda_s h_s(c)=0$, where
the eigenvalue $\lambda_s =0$ at $s=a^*$, and where the tail is
given by the R-eigenfunction, $h_{a^*}(c) \sim c^{-a^* -d-\nu}$.

The universal feature of thermodynamic equilibrium in systems with
{\it conservative} interactions is the Gibbs' state in which the
v.d.f. is  always Maxwellian, $\exp[-c^2]$. The conclusion from
our analysis tends to be that the {\it only generic} feature of
the present scaling solutions for systems with dissipative
interactions is overpopulation of the high energy tails relative
to Maxwellians. These over-populated tails come in two shapes,
either stretched exponentials, $ f(c) \sim exp[- \beta c^b]$ (see
Fig.\ref{fig3.4} and Fig.\ref{fig3.6}), or power law tails, $f(c)
\sim c^{-s}$ with $s=a+d+\nu$ (see Fig.\ref{fig4.4} and
Fig.\ref{fig4.7}). The stretching exponent $b= 1+\half \nu$, and
the power law exponent $s=a+d+\nu$ with $a=a_d (\alpha)$ depend
linearly on the interaction exponent $\nu$, and they depend {\it
sensitively} on the inelasticity and on the driving device used
\cite{ETB-II}.

Furthermore, to test the theoretical predictions about tails in
the  velocity distributions by laboratory experiments, there is
the additional problem that the fundamental inter-particle
interactions, as well as the interactions between granular
particles and macroscopic driving devices are not really known.

\setcounter{equation}{0}
\appendix
\renewcommand{\theequation}{A.\arabic{equation}}
\setcounter{equation}{0}
\section{ Boltzmann collision operator}
\label{app:A}

\subsection{ Nonlinear operator $I$}

A convenient representation of the nonlinear collision operator is
obtained by changing  integration variables $\bw =
(\bw_\bot,w_\parallel)$ in
\Ref{BE} into $(\bu_\bot, u_\parallel)=
(\bw_\bot^{**},w_\parallel^{**})$, where $\bw^{**}=\bw^*(1/\alpha)$
according to \Ref{coll}. This implies
\ba
\bv_\bot^{**} = \bv_\bot, \quad
v^{**}_\parallel=\frac{1}{\alpha}(-q v_\parallel +p w_\parallel)
\nonumber \\
\Label{A1}
\bu_\bot=\bw_\bot^{**}=\bw_\bot, \quad
u_\parallel = w^{**}_\parallel =
\frac{1}{\alpha}(p v_\parallel -q w_\parallel),
\ea
where $p=1-q=\frac{1}{2}(1+\alpha)$. Consequently $d\bw =
(\alpha/q)d\bu$ and $v_\parallel-w_\parallel=(\alpha/q)(u_\parallel
- v_\parallel)$. Inserting these results in \Ref{BE} gives the
following representation of the nonlinear collision term,
\be\Label{A2}
I(v | F) = \int_\bn \int d\bu
\left[
\frac{1}{q} K\left(g_\bot, \frac{g_\parallel}{q}\right)
F\left(\bv_\bot + \bn \frac{v_\parallel-p u_\parallel}{q}\right)
-K(g_\bot, g_\parallel)F(v)
\right] F(u)
\ee
with $\bg=\bu-\bv$. By substituting $F(v)=\delta(\bv)$ and using
the explicit form of the collision rate, $K=g^\nu
|\hat{g}_\parallel|^\sigma =g^{\nu-\sigma}|{g}_\parallel|^\sigma
$, it is easy to verify that $I(v|\delta)=0$, at least for $\nu
\ge 0$. As we have seen in Section II, {\it negative} $\nu-$values
correspond in the case of elastic interactions to a potential
$V(r) \sim 1/r^n$ with exponent $ 0 < n \leq 2(d-1)$, representing
weak interactions. To fix the divergence difficulties in $K$ at
$g=0$ we introduce for $\nu <0$ a small$-\rg$ {\it cut off} in the
collision frequency, i.e. $K \propto (g + v_0(t)\Delta)^\nu $,
where $\Delta$ is a positive constant of order unity. All
applications in the present paper of the cut off $K$ for $\nu <0$
concern  NESS's for driven cases, where $v_0(\infty) $ is {\it
finite} and non-vanishing. So $  v_0 \Delta$ may be simply
replaced by $\Delta$. The reason for adding the factor $v_0$ is
solely for theoretical convenience, and ensures that the Boltzmann
equation after rescaling leads to time-independent scaling
equations for models with $\nu <0$. For asymptotically large
velocities, $ \bv = v_0 \bc $ with $c\gg1$, the $w-$integration in
$I(v|F)$ is typically restricted to $w \leq v_0$, and the
collision frequency becomes $K \sim (v+  v_0 \Delta)^{-|\nu|}
|\hat{v}_\parallel|^\sigma$. Consequently, the collision term with
the cut off collision frequency also satisfies the relation,
$I(v|\delta) =0$, because the product $K({v}_\bot,v_\parallel)
\delta (\bv) $ is well-behaved, the gain and loss term cancel, and
the linearized collision operator $\Lambda$ in \Ref{31} is well
behaved for $\nu <0$ as well. Moreover, for asymptotically
large$-v$ the collision operator $I$ and $\Lambda$ are independent
of the cut off. We also note that for $ \nu <0$ the collision
frequency $K$ at large $v$ is vanishing like $1/v^{|\nu|}$, and
the mechanisms for randomizing the large$-v$ particles and
redistributing their energy over all particles is essentially
lacking. Consequently the tail distribution is expected to be
heavily overpopulated.

In our DSMC  method for numerically solving the nonlinear Boltzmann
equation for {\it negative} $\nu-$values, a similar small velocity
cut-off in the collision rate is  required as well.

%%%%%%%%%%%%%%%%%%%%%%%%%%%%%%%%%%%%%%%%%%%%%%
\subsection{Evaluation of eigenvalue $\lambda_s(\sigma)$ }
\label{appA2}

The evaluation of the eigenvalue starts from expression \Ref{38}
containing four terms. The first, second and fourth one are
determined by the angular average of $|\cos
\theta|^s$, i.e.
\be\Label{A5}
\beta_s = \int_\bn |\hat{\bf a}_\parallel|^s =
\frac{\int^{\pi/2}_0  d\theta (\sin \theta)^{d-2} |\cos \theta|^s}{
\int^{\pi/2}_0  d\theta (\sin \theta)^{d-2}}
=\displaystyle
\frac{\Gamma(\frac{s+1}{2}) \Gamma(\frac{d}{2})}
       {\Gamma(\frac{s+d}{2})\Gamma(\frac{1}{2}) },
\ee
where $\hat{ {\bf a}}$ is a fixed direction. This integral
converges for $\sigma >-1$. The third term, denoted by
$\lambda^{(a)}_s$, is more complicated, i.e.
\ba \Label{A6}
\lambda_s^{(a)} & = & \int_\bn |\hat{c}_\parallel|^\sigma \,
|\hat{\bc}_\bot +q \hat{c}_\parallel \bn |^s
\,=\,\int_\bn |\hat{c}_\parallel|^\sigma \,
(\hat{\bc}_\bot^2 +q^2 \hat{c}_\parallel^2)^{s/2}
\nn
&=& -\frac{
\int_0^{\pi/2} d\theta \,(\sin\theta)^{d-1}
\left[1-z \cos^2\theta
\right]^{s/2} (\cos\theta)^\sigma }
{\int_0^{\pi/2} \,d\theta\,(\sin\theta)^{d-2}}
\ea
In the last equality we have used the relation $\hat {\bf
c}_\bot^2 +\hat c_\parallel^2=1$ and defined $z=1-q^2$. Next, we
introduced the change of variables $\mu=\hat c_\parallel^2 =
\cos^2\theta$ and $d\mu = -2 \,\cos\theta\,\sin\theta\,d\theta$,
to obtain
\ba
\Label{A7}
\frac{\lambda_s^{(a)}}{\beta_\sigma} &=&
-\,\frac{1}{B(\frac{d-1}{2},\half)} \, \int_0^1 d\mu\,
\mu^{(\sigma-1)/2} (1-\mu)^{(d-3)/2} (1-z\mu)^{s/2}
\nn
&=& - \,_2F_1\left(-\frac{s}{2},\frac{\sigma+1}{2};
\frac{\sigma+d}{2}\biggl|\,z \right)
\ea
where $B(x,y) = \Gamma(x)\Gamma(y)/\Gamma(x+y)$. The
integral can be identified as a representation of the
hyper-geometric function \cite{Grads}
\ba
\Label{A8}
_2F_1(a,b;c|z) &=& \sum_{n=0}^\infty \frac{(a)_n \,(b)_n \, z^n}{n!
(c)_n}
\nn &=& ({1/B(b,c-b)}){\int_0^1 dx \, x^{b-1}\,
(1-x)^{c-b-1}\, (1-zx)^{-a}},
\ea
where $(a)_n=\Gamma(a+n)/\Gamma(a)$ and $\beta_\sigma$ is given by
\Ref{A5}. Combining results yields the eigenvalue
\be
\Label{A9}
\lambda_s (\sigma)\,=\, \displaystyle\beta_\sigma \left\{ 1+\delta_{s,0} -
 _2\!F_1\,\left(\textstyle{-\frac{s}{2},
\frac{\sigma+1}{2};\frac{\sigma +d}{2}} \,|\,1-q^2\right)\right\} -p^s
\beta_{s+\sigma}.
\ee

%%%%%%%%%%%%%%%%%%%%%%%%%%%%%%%%%%%%%%%%%%%%%%%%%%
\subsection{Linear operator $\Lambda$ and its eigenfunctions }
\label{appA3}

To construct $\Lambda$,  defined in \Ref{31}, we start from the
representation $I(c|\delta)$ in \Ref{A2} with $F(c) = \delta({\bf
c}) +h(c)$, and use $I(c|\delta)=0$. This yields after some
transformations,
\be
\Label{A12}
\Lambda \,=\, \Lambda^{(a)} + \Lambda^{(b)} + \Lambda^{(l)},
\ee
where
\ba \Label{A13}
\Lambda^{(l)} h(c) &=&  \int_\bn \,K({\bf c}_\perp,c_\parallel) h(c)
+ \delta({\bf c}) \int_\bn\int d{\bf u}\, K({u}_\perp,u_\parallel)
h(u) \nn
\Lambda^{(a)} h(c) &=& - \int_\bn \frac{1}{q}\,K \left( c_\perp,
\frac{c_\parallel}{q}\right) h\left(\biggl|{\bf c}_\perp +{\bf n}
\frac{c_\parallel}{q}\biggl|\right) \nn
\Lambda^{(b)} h(c) &=& -\int_\bn \frac{1}{p}\,K\left( u_\perp,
\frac{c_\parallel -u_\parallel}{p}\right) h\left(\biggl|{\bf u}_\perp +{\bf n}
\frac{c_\parallel-q u_\parallel}{p}\biggl|\right)
\delta({\bf c}_\perp+ \bn\, u_\parallel)
\nn
&=& -\int_\bn\int d {\bf u}\, K(u_\perp,u_\parallel) \,h(u)
\delta({\bf c}-\bn\, p\, u_\parallel).
\ea
Inspection of the expressions for $\Lambda^{(a)}$ and
$\Lambda^{(b)}$ shows that $\Lambda c^r =\mu_r c^{r+\nu}$, where
the eigenvalue $\mu_r$ is different form $\lambda_r$ in
\Ref{A9}.   Rather than first calculating $\mu_r$ explicitly, and
then determining $r$ such that $\mu_r= \lambda_s$, we simply state
that $r=-s -d +2$, i.e.
\be \Label{A14} \Lambda c^{-s-d-\nu} \,=\, \lambda_s c^{-s-d} \qquad
(s>0,c>0),
\ee
and verify {\it a posteriori} that $\lambda_s$ is equal to the
expression \Ref{A9}.

Consider first $\Lambda^{(a)}$ in \Ref{A13}, and substitute $h(c) =
c^{-s-d-\nu}$. By performing similar substitutions as in the steps
\Ref{A5} to \Ref{A7} we obtain
\ba \Label{A15}
\lambda_s^{(a)} &=& -q^{-\sigma-1} \int_\bn |\hat c_\parallel|^\sigma\,
\left[1+\left(\frac{1}{q^2}-1
\right) \hat c_\parallel^2
\right]^{-(s+d+\sigma)/2} \nn
&=& -\beta_\sigma q^{-\sigma-1} \,\, _2F_1\left(
\frac{s+d+\sigma}{2},\frac{\sigma+1}{2};\frac{\sigma+d}{2}\biggl|\frac{z}{z-1}
\right) \nn
&=& \beta_\sigma \,\, _2F_1\left(
-\frac{s}{2},\frac{\sigma+1}{2};\frac{\sigma+d}{2}\biggl|1-q^2
\right),
\ea
where $z=1-q^2$. The last line has been obtained with the help of
Eq. (9.131.1) of Ref. \cite{Grads}, and is identical to the term
containing the function $_2F_1$ in \Ref{A9}. To evaluate
$\Lambda^{(b)} h(c)$ we use the relation \Ref{A13},
\be \Label{A16}
\delta({\bf c} -\bn p \,u_\parallel) \,=\,
\frac{1}{p c^{d-1}} \, \delta(\hat{\bf c} -\bn) \,
\delta \left(|u_\parallel| -\frac{c}{p}\right),
\ee
and integrate out both delta functions with the result,
\be
\Label{A17}
\Lambda_s^{(b)} h(c) \,=\, -\frac{2\, c^{\sigma+1-d}}{\Omega_d\, p^{\sigma+1}}\,
\int d{\bf u}_\perp\, [u^{\nu-\sigma} h(u)]_{|u_\parallel|=c/p}.
\ee
Setting $h(c) = 1/c^{s+d+\nu}$ and changing variables ${\bf
u}_\perp = {\bf x}\, c/p$, we obtain
\be
\Label{A18}
\lambda_s^{(b)} \,=\, -\frac{2\,p^s}{\Omega_d} \,\int \frac{d^{d-1} {\bf x}}{
(x^2+1)^\alpha} \, = \,- p^s \,\beta_{s+\sigma}.
\ee
Here we used the relations \Ref{A5} and $\alpha = (s+d+\sigma)/2$
in agreement with the last term in \Ref{A9}. One similarly verifies
that $\Lambda^{(l)}$ in \Ref{A12} for $c>0$ yields
\be
\Label{A19}
\Lambda^{(l)} \frac{1}{c^{s+d+\nu}}=  \int_\bn
\frac{K({\bf c}_\perp,c_\parallel)}{c^{s+d+\nu} }
\,=\,  \, \int_\bn \frac{|\hat c_\parallel|^\sigma}{c^{s+d}} =
\frac{\beta_\sigma}{c^{s+d}}.
\ee
In summary, we have obtained for $c>0,s>0$,
\be
\Label{A20}
\Lambda c^{-s-d-\nu} \,=\, \lambda_s(\sigma) \,c^{-s-d}
\qquad \hbox{and} \qquad
\Lambda^\dagger c^s \,=\, \lambda_s(\sigma) \, c^{s+\nu}
\ee
and for $s=0$ one  can verify that,
\be
\Lambda \,\delta({\bf c}) \,=\, 0 \qquad \hbox{and} \qquad
\lambda^\dagger 1\,=\,0 \quad (\Lambda_0(\sigma)=0).
\ee

\renewcommand{\theequation}{B.\arabic{equation}}
\setcounter{equation}{0}
\section{ Estimates of $ I $ for
exponentially bounded functions}
\label{app:B}

The goal of this appendix  is to improve the large$-c$ estimate of
the nonlinear collision operator, $I(c|f) \simeq - \beta_\sigma
c^\nu f$, as given in the literature \cite{vNE-granmat,EB-Rap}.
The basic idea of the method is to split the v.d.f. $f(c) = f_0(c)
+ h(c)$ into a regular {\it bulk} part $f_0(c)$, say of Gaussian
shape, and a {\it small} singular part $h(c)$. The bulk part
carries the mass of the distribution, i.e. $\int d\bc f_0(c) =1$.
Then $f_0(c)$ vanishes effectively beyond the thermal range, say
$c \gtrsim 3$, where the small singular part $h(c)$ is living (see
DSMC results in Fig.\ref{fig3.5} for $d=1$, and Fig.\ref{fig3.6}
for $d=2$). For $c\gg 1$ the bulk part can be viewed  in zeroth
approximation as a normalized Dirac delta function $\delta (\bc)$,
and we represent the v.d.f. as $f(c) = \delta (\bc) +h(c)$.
Moreover, this caricature of the bulk part, $\delta (\bc)$, makes
the nonlinear collision term vanish according to \Ref{29}, and we
can {\it linearize} the large$-c$ form of $I (c|\delta +h) =
-\Lambda h(c)$.  Inspection of the contributions
$\Lambda^{(l)},\Lambda^{(a)}$, and $\Lambda^{(b)}$ in
\Ref{A12}-\Ref{A13} in Appendix A  shows that the term in
$\Lambda^{(l)} h$, proportional to $\delta(\bc)$, and
$\Lambda^{(b)} h \propto \int_\bn \delta(\bc_\bot)$, are short
range, and consequently the collision term for $c\gg 1$ reduces to
its asymptotic form,
\ba \Label{B1}
&I_\infty  (c|\delta+h) =  -\Lambda_\infty h (c)&
\nn &=
- K({c}_\bot, {c_\parallel}) h({\bc}) + (1/q) \int_\bn K({c}_\bot,
{c_\parallel}/{q}) h(|{\bc}_\bot +\bn {c_\parallel}/{q}|)&,
\ea
where $\int_\bn K({c}_\bot, {c_\parallel}) =c^\nu
\beta_\sigma$, as  implied by \Ref{A19}.  We need an asymptotic
estimate of the second term on the r.h.s. of
\Ref{B1}, $\Lambda^{(a)}$ , when acting on exponentially bounded functions $h(c)$ of
the form \Ref{315},
\be \Label{B2}
\Lambda^{(a)} f \sim \Lambda^{(a)} c^\chi \exp[ - \beta c^b +
\beta' c^{b'}] \equiv {\cal J}(c) c^\nu f(c)
\ee
with $b>0$. Here the last term denotes the outcome of the
calculations. To calculate ${\cal J}(c)$ we introduce the
variables,  $ \hat{c}^2_\parallel = 1-\hat{\bc}^2_\bot =\cos^2
\theta \equiv \mu$, and  $\zeta =(1-q^2)/q^2$. Then the integrand
in \Ref{B1} is a function $H(|\hat{c}_\parallel |) = K f$  of
$|\hat{c}_\parallel|$ with,
\ba \Label{B3}
 |{\bc}_\bot  +\bn \,{c}_\parallel/q| &=&[c^2_\bot
+{c}^2_\parallel/ q^2]^{1/2} = c \,[1+\zeta \mu]^{1/2}
\nn  K({\bc}_\bot,{c}_\parallel/q) &=&\left| {{c}_\parallel}/{q}
\right|^\sigma [c^2_\bot + {{c}^2_\parallel}/{q^2}]^{(\nu-\sigma)/2} =
q^{-\sigma} c^\nu \mu^{\sigma/2} [1 +\zeta\mu]^{(\nu-\sigma)/2}
\nn  f(|{\bc}_\bot + \bn {c_\parallel}/{q}|) &=&
c^\chi (1+\zeta\mu)^{\chi/2} \exp [ -\beta c^b (1+\zeta\mu)^{b/2}]
\nn [2mm] \int_\bn H(|\hat{c}_\parallel |) &=&
\frac{\int^{\pi/2}_0 d \theta (\sin \theta)^{d-2} H(\cos
\theta) }{\int^{\pi/2}_0 d \theta (\sin \theta)^{d-2}} =
\frac{\int^1_0 d\mu (1-\mu)^{(d-3)/2} \mu^{-1/2}
H(\sqrt{\mu})}{B(\frac{d-1}{2},\frac{1}{2})},
\ea
where $B(x,y) = \Gamma(x)\Gamma(y)/\Gamma (x+y)$. With  the help of
\Ref{B1}-\Ref{B3} we obtain,
\be \Label{B4}
{\cal J}(c) = \frac{1}{q^{\sigma+1} B(\frac{d-1}{2}, \frac{1}{2})}
\int^1_0 d \mu (1-\mu)^{(d-3)/2} \mu^{(\sigma-1)/2}
(1+\zeta\mu)^{(\nu-\sigma+\chi)/2} E(\mu)],
\ee
where $E(\mu)$ is defined as,
\ba \Label{B4a}
E(\mu)& =& \exp\left\{ -\beta c^b [(1+\zeta\mu)^{b/2} -1] + \beta'`
c^{b'} [(1+\zeta\mu)^{b'/2} -1]  \right\}
\nn &\simeq& \exp[ -X \mu +X'\mu]   \qquad (\mu \simeq 0)
\ea
with $X = \half \beta b \zeta c^b$ and $X' = \half \beta b' \zeta
c^{b'}$. For $c\gg 1$ the integrand in \Ref{B4} vanishes
exponentially fast, unless $\mu = \cos^2 \theta \simeq 0$ (grazing
collisions). Near $\mu=0$ factors of the form $(1+A\mu)^a$  can be
approximated  by unity, and  $E(\mu)$ itself  by the second line of
\Ref{B4a}. Changing integration variables $X \mu =t $, and taking
the large$-c$ or large$-X$ limit yields for the integral,
\be
\int^1_0 d\mu\mu^{(\sigma-1)/2} e^{-X\mu +X' \mu} = \Gamma ((\sigma+1)/2)/
X^{(\sigma+1)/2} \{ 1 +{\cal O}(1/X)\}.
\ee
Consequently the large$-c$ behavior of ${\cal J}(c)$ is,
\be \Label{B5}
{\cal J}(c) \sim \frac{\Gamma (\frac{\sigma+1}{2})} {q^{\sigma+1}
B(\frac{d-1}{2}, \frac{1}{2})} \left(\frac{2}{\beta b \zeta c^b}
\right)^{(\sigma+1)/2} \equiv \beta_\sigma {\cal K}_\sigma c^{- b
(\sigma+1)/2}.
\ee
Inserting $\beta_\sigma$ from \Ref{A5} and $\zeta =(1-q^2)/q^2$ we
obtain for the coefficient,
\be \Label{B6}
{\cal K}_\sigma =\frac{\Gamma (\frac{\sigma+1}{2})}{\beta_\sigma
B(\frac{d-1}{2}, \frac{1}{2})} \left(\frac{2}{\beta b (1-q^2) }
\right)^{(\sigma+1)/2} = \frac{\Gamma (\frac{d+\sigma}{2})}{\Gamma
(\frac{d-1}{2})} \left(\frac{2}{\beta b (1-q^2)}
\right)^{(\sigma+1)/2}.
\ee
By combining \Ref{B1}, \Ref{B2} and \Ref{B5}, and observing that
$f(c) \sim h(c)$ for $c \gg 1$,  the nonlinear collision term takes
the asymptotic form,
\be \Label{B7}
I_\infty (c|f)= -\beta_\sigma c^\nu [1 -{\cal K}_\sigma c^{-a}]
f(c),
\ee
with $ a = \half b(\sigma+1)$. The derivation also shows that for
$d=1$ the algebraic correction term in \Ref{B6} is absent because
the angular integral $\int_\bn$ is missing. So for $|c|
> 1 $ we have in one dimension,
\be \Label{B8}
\Lambda h(c) = |c|^\nu \{h(c) -q^{-\nu -1}h(c/q)-p^{-\nu
-1}h(c/p)\}.
\ee
Here $q=1-p= \half (1-\alpha)$ and $0 <q \leq \half p<1$. So, as long
as $\alpha <1$, the contributions from $h(c/q)$ and $h(c/p)$,
originating from the gain term, are exponentially separated from the
loss term in case $h(c) \sim \exp[-\beta c^b] $ with $b>0$.

\section*{Acknowledgments}
It is a pleasure to dedicate this article to Carlo Cercignani, on
the occasion of his 65-th birthday,  in honor of his great
contributions to kinetic theory in general, and to the study of
inelastic Maxwell molecules in particular. We also acknowledge
useful discussions with Ricardo Brito, Eli Ben-Naim and Paul
Krapivsky. M.H.E.  is supported by Secretar\'\i a de Estado de
Educaci\'on y Universidades (Spain) and the research project
FIS2004-271. E.T. thanks the European Community's Human Potential
Program under contract HPRN-CT-2002-00307 (DYGLAGEMEN).

\vspace{25mm}

%Figures Section II

\newpage
\renewcommand{\thefigure}{3.\arabic{figure}}
\setcounter{figure}{0}

\begin{figure}
$$\includegraphics[angle=0,width=.6 \columnwidth]{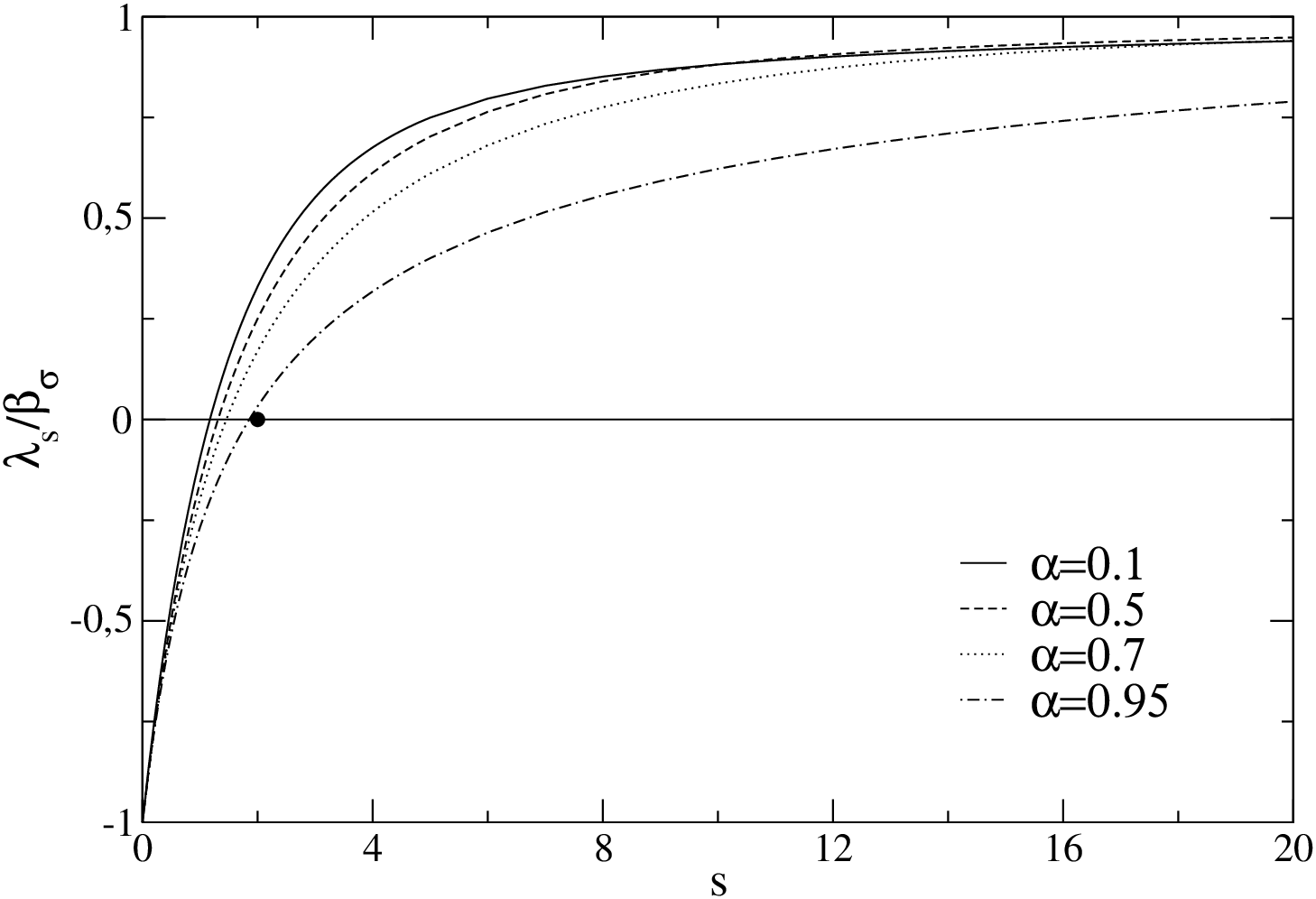}$$
\vskip 2mm
\caption{
%(see file "lambda-s.eps", date 25/6).
Eigenvalue spectrum $\lambda_s(\sigma)$ for $s \geq 0$ of the
collision operator in the inelastic soft sphere models
$(\sigma,\nu)$, and shown for various values of the coefficient of
restitution $\alpha$. The ordinate shows
$\lambda_s(\sigma)/\beta_\sigma$ for $d=2,\sigma=\nu=1$, which
approaches 1 for $s \to \infty$, and $-1$ for $s \to 0^+$, whereas
$\lambda_0(\sigma)=0$. The bullet is centered at $(2,0)$ showing
that  $a^* <2$, where $\lambda_{a^*}=0$ (intersection point with
s-axis), and  also that $s^* \to 2$ as $\alpha \to 1$ (energy
conservation in elastic case).} \Label{fig2.1}
\end{figure}

%Figures Section III

\renewcommand{\thefigure}{5.\arabic{figure}}
\setcounter{figure}{0}

\begin{figure}
\null\vskip 1mm
$$\includegraphics[angle=0,width=.6 \columnwidth]{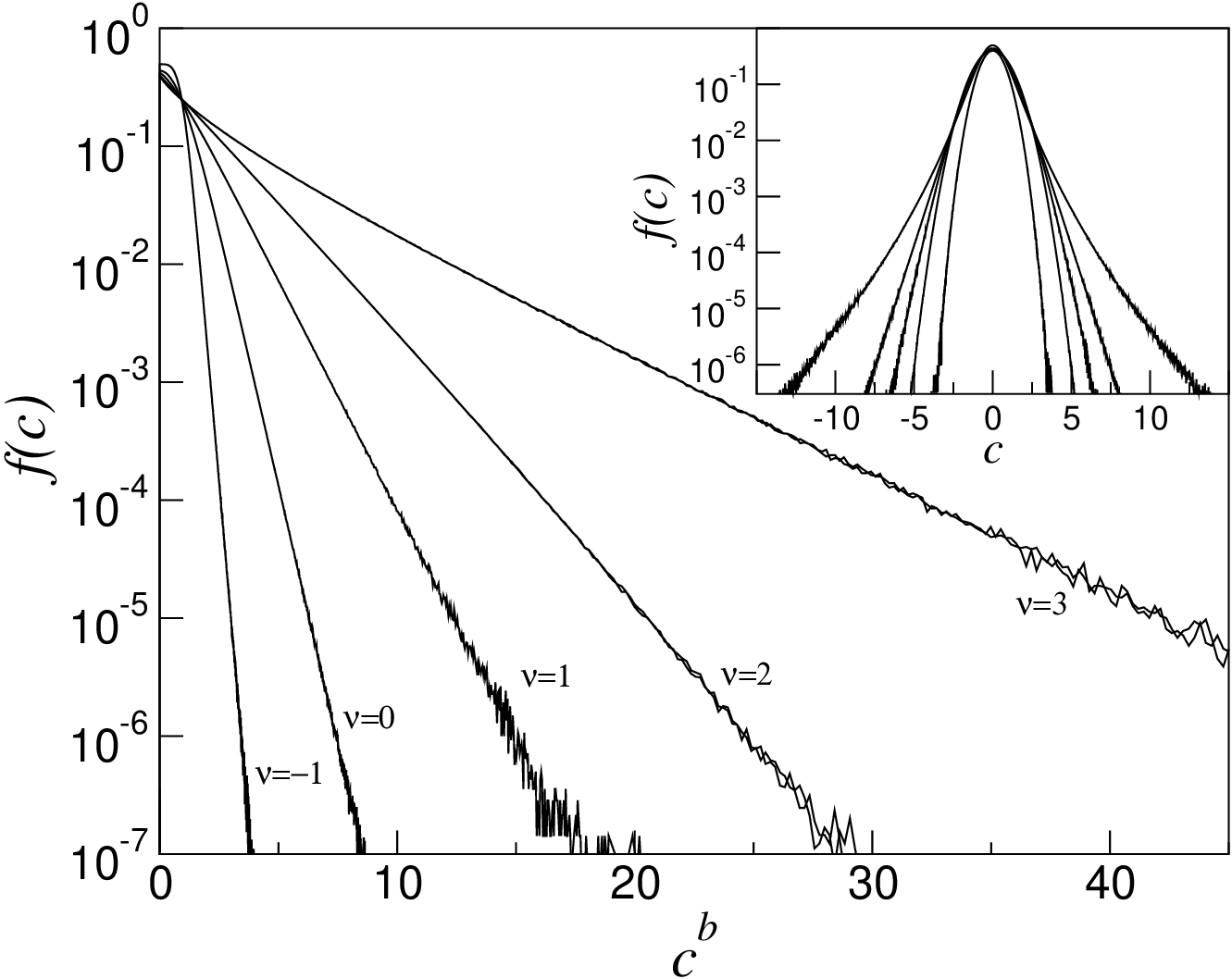}$$
\caption{ The figure corresponds to a stable non equilibrium
steady state at $\alpha=0$ and
$\nu= \{-1,0,1,2,3\}$ for
{\it white noise-driven} 1-D systems. The DSMC data show
that $\ln f(c)$ is linear in $c^b$ with $b=1+\nu/2>0$.
 The insert with $\ln f(c)$ versus $c$ shows the highest
over-population at $\nu =-1$ or $b=\half$ (outercurve), whereas
the curve for $b=\nu=2$ is a Gaussian. Underpopulation occurs
for $\nu >2$.\\ }
\Label{fig3.4}
\end{figure}

\begin{figure}
$$\includegraphics[angle=0,width=.6 \columnwidth]{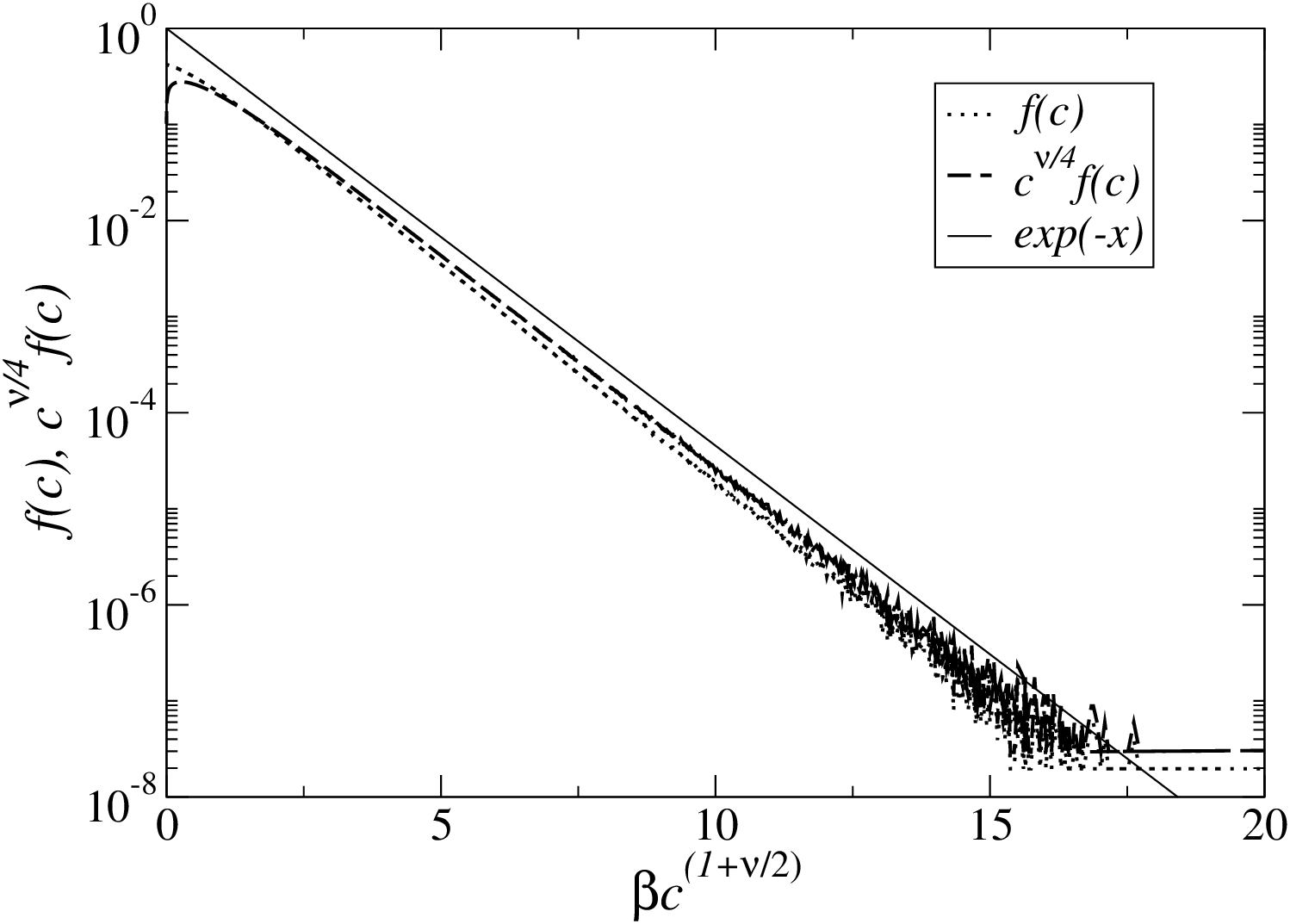}$$
\caption{Case of a white-noise driven 1-D system, with
$\alpha=0$ and $\sigma=\nu=1$. The dotted line displays
$f(c)$ vs. $\beta c^b$ while the dashed line corresponds to
$c^{\nu/4} f(c)$ versus $\beta c^b$  ($b=1 +\nu/2$). Taking into
account the sub-leading correction $c^{\nu/4}$ yields a better
agreement with the theoretical result, however the figure
clearly shows that this correction is very small.
}
\Label{fig3.5}
\end{figure}

\begin{figure}
$$\includegraphics[angle=0,width=.6 \columnwidth]{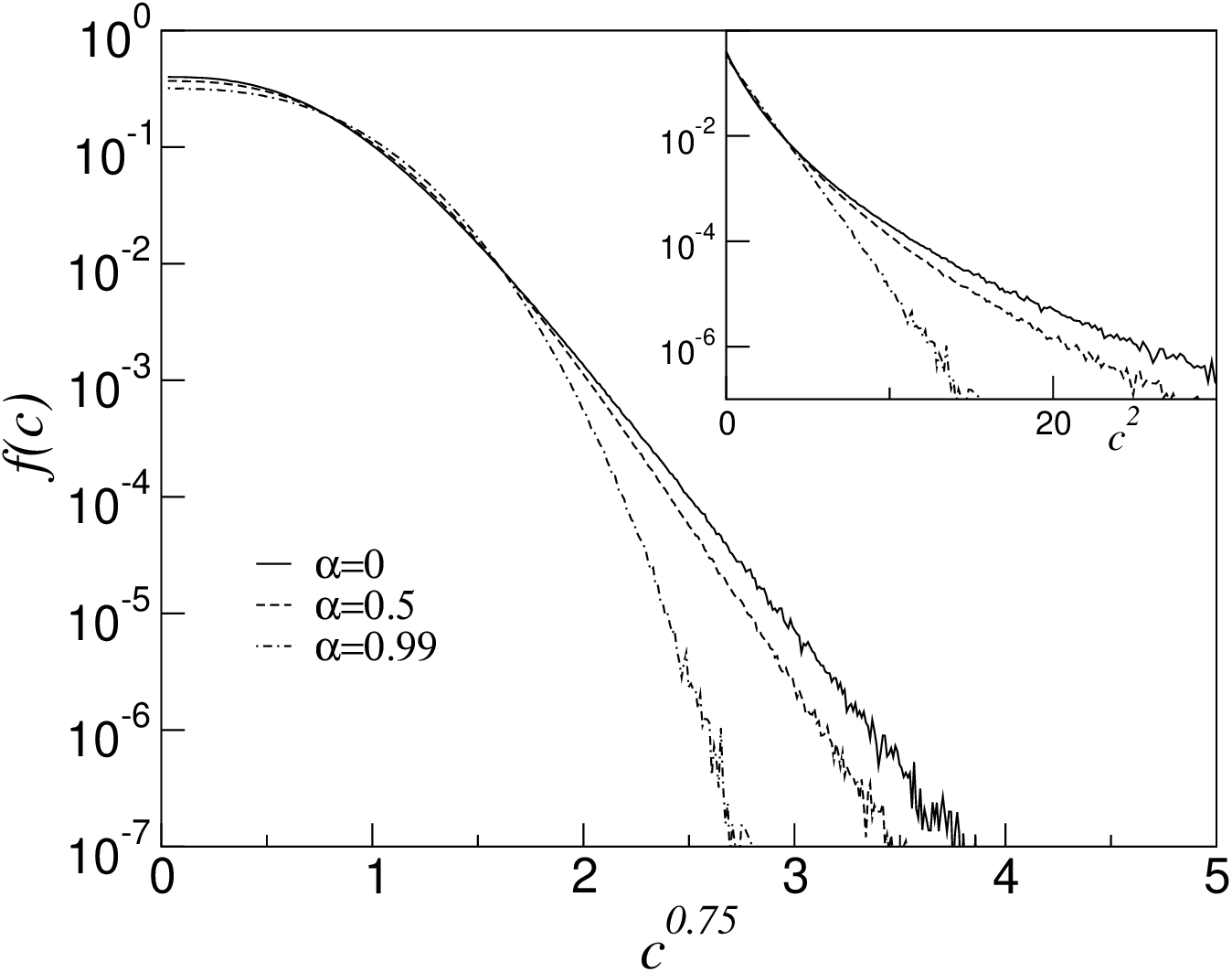}$$
\caption{Case of a white-noise driven 2-D
soft sphere model $\sigma=\nu= - 1/2$. The
DSMC data for $\ln f(c)$ vs $c^b$ with $b=1+\nu/2=3/4$,
confirm the stretched Gaussian tails $\exp[-\beta
c^b]$, where $\beta$
%in \Ref{R10}
increases with $\alpha$. The
insert with $\ln f(c)$ vs $c^2$
shows the overpopulation in the tail.} \Label{fig3.6}
\end{figure}

\begin{figure}
$$\includegraphics[angle=0,width=.6 \columnwidth]{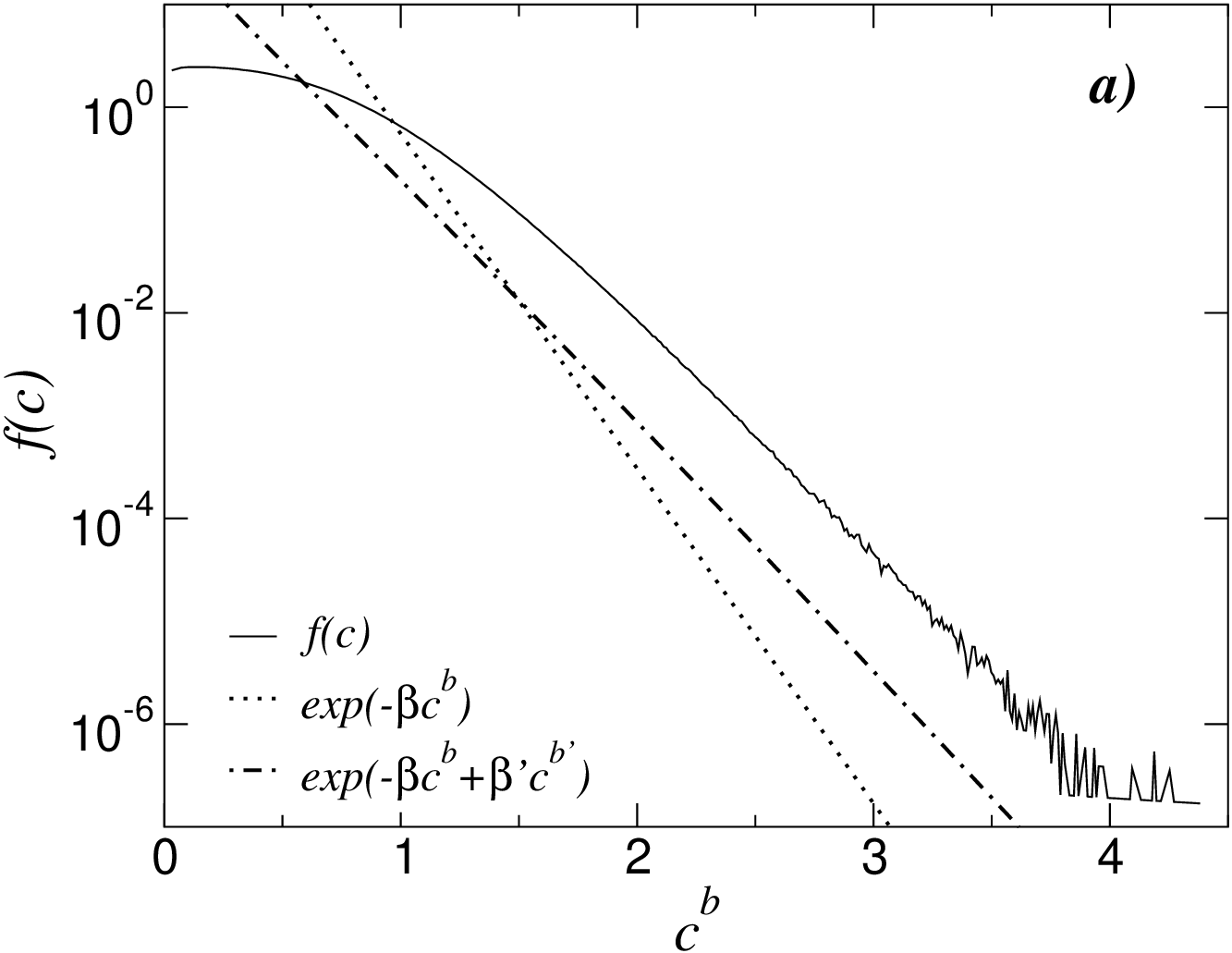}$$
\hskip 2mm
$$\includegraphics[angle=0,width=.6 \columnwidth]{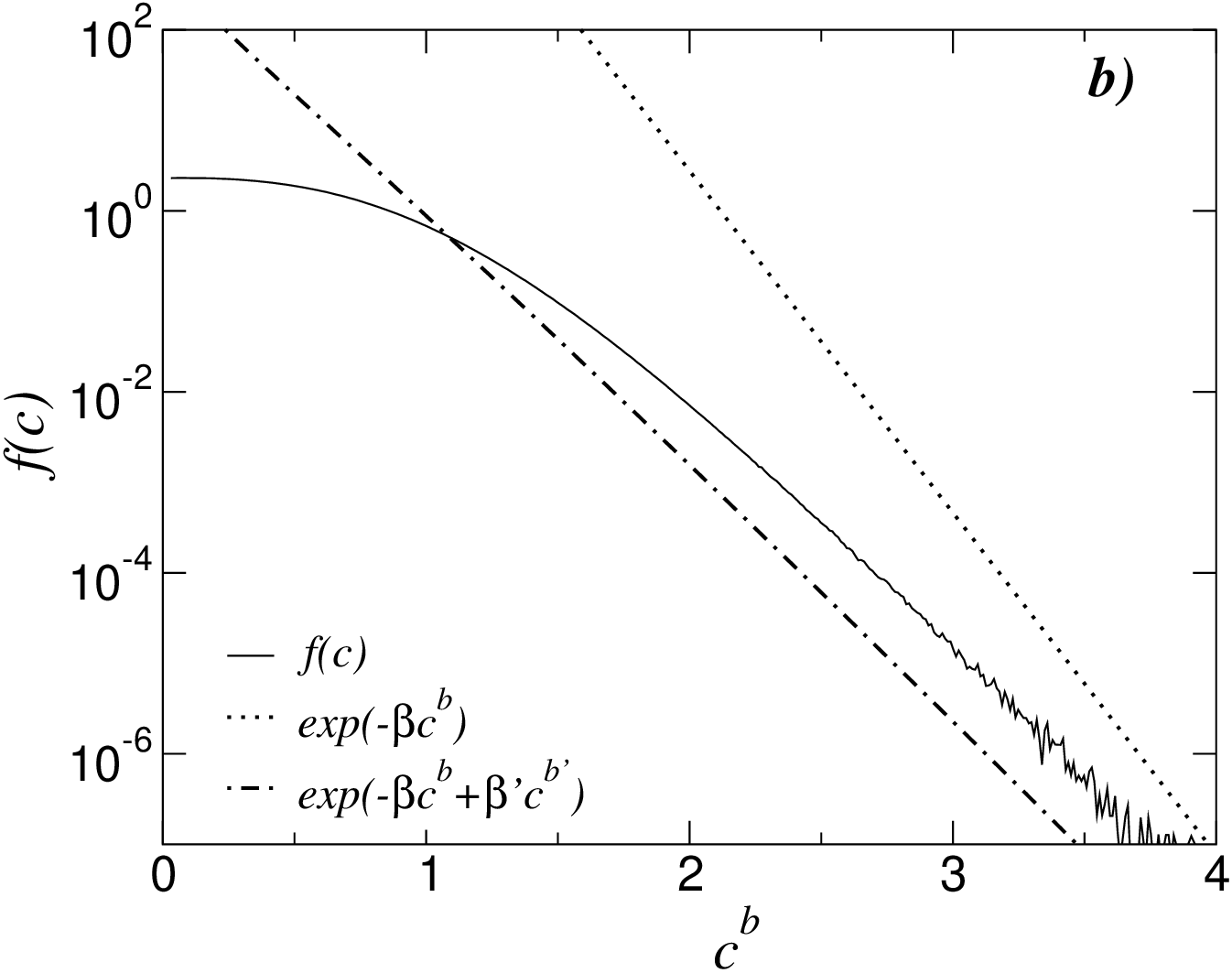}$$
\caption{Illustration of the relevance of the sub-leading
correction (white noise, $d=2,\sigma=\nu=-1/2$). Graph a)
corresponds to $\alpha=0$, and graph b) to $\alpha=1/2$. The
velocity distribution function is shown as a function of $c^b$,
together with two asymptotic expressions. The first one is the
first order prediction $\exp(-\beta c^b)$, with $\beta$ calculated
by DSMC and $b=1+\nu/2=3/4$ and yields a very bad agreement.
Inclusion of the sub-leading correction $\exp(-\beta c^b +\beta'
c^{b'})$ (still plotted vs. $c^b$) with $\beta'$  calculated by
DSMC, and $b'=9/16$, gives a very good agreement. Very striking is
the fact that in such a plot, $f(c)$ vs $c^b$ produces a linear
high energy tail (in spite of the importance of the sub-leading
correction), which would then be well fitted with an effective
value of $\beta$: $f(c)\sim \exp(-\beta_{\text{eff}}\, c^b)$. As
shown here, such an effective value is markedly different from the
true $\beta$ in \Ref{55}, which indicates that any fitting
procedure, aiming at computing $\beta$, is doomed to fail. }
\Label{fig3.6ab}
\end{figure}

% figures section IV

\renewcommand{\thefigure}{6.\arabic{figure}}
\setcounter{figure}{0}

\begin{figure}
$$\includegraphics[angle=0,width=.6 \columnwidth]{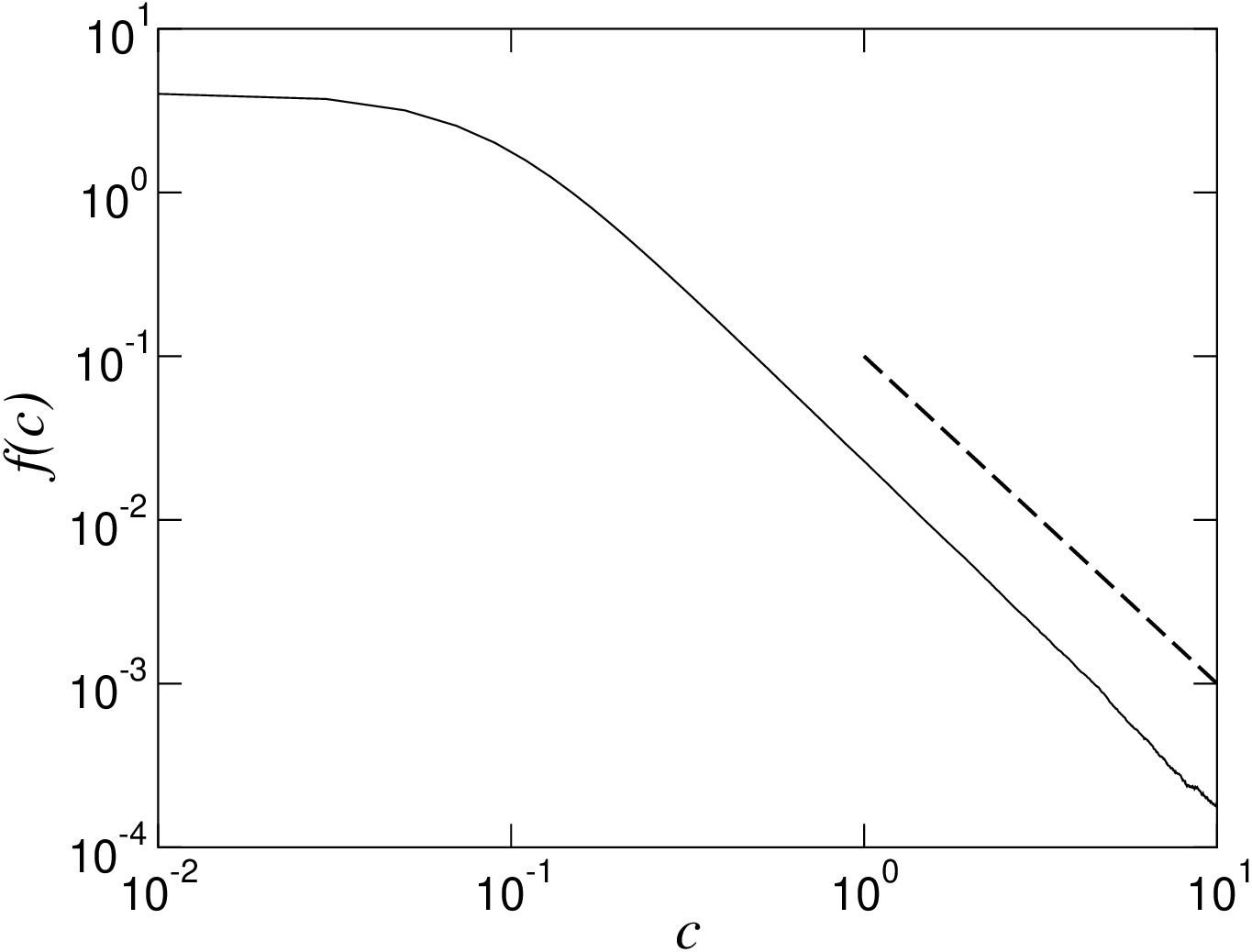}$$
\caption{
%( file "driven$\_$nu-2.eps of 27/2);WN:($d=1$).
WN-driven system at stability threshold ($b=1+\nu/2$=0) in 1-D. The DSMC
data show a power law tail in the 1-D soft sphere model ($\nu=
-2$). It is compared with the analytic prediction $f(c)\sim
1/c^{a+d+\nu} \sim 1/c^2$ ($a=3$), shown by the dashed line.} \Label{fig4.4}
\end{figure}

\begin{figure}
$$\includegraphics[angle=0,width=.6 \columnwidth]{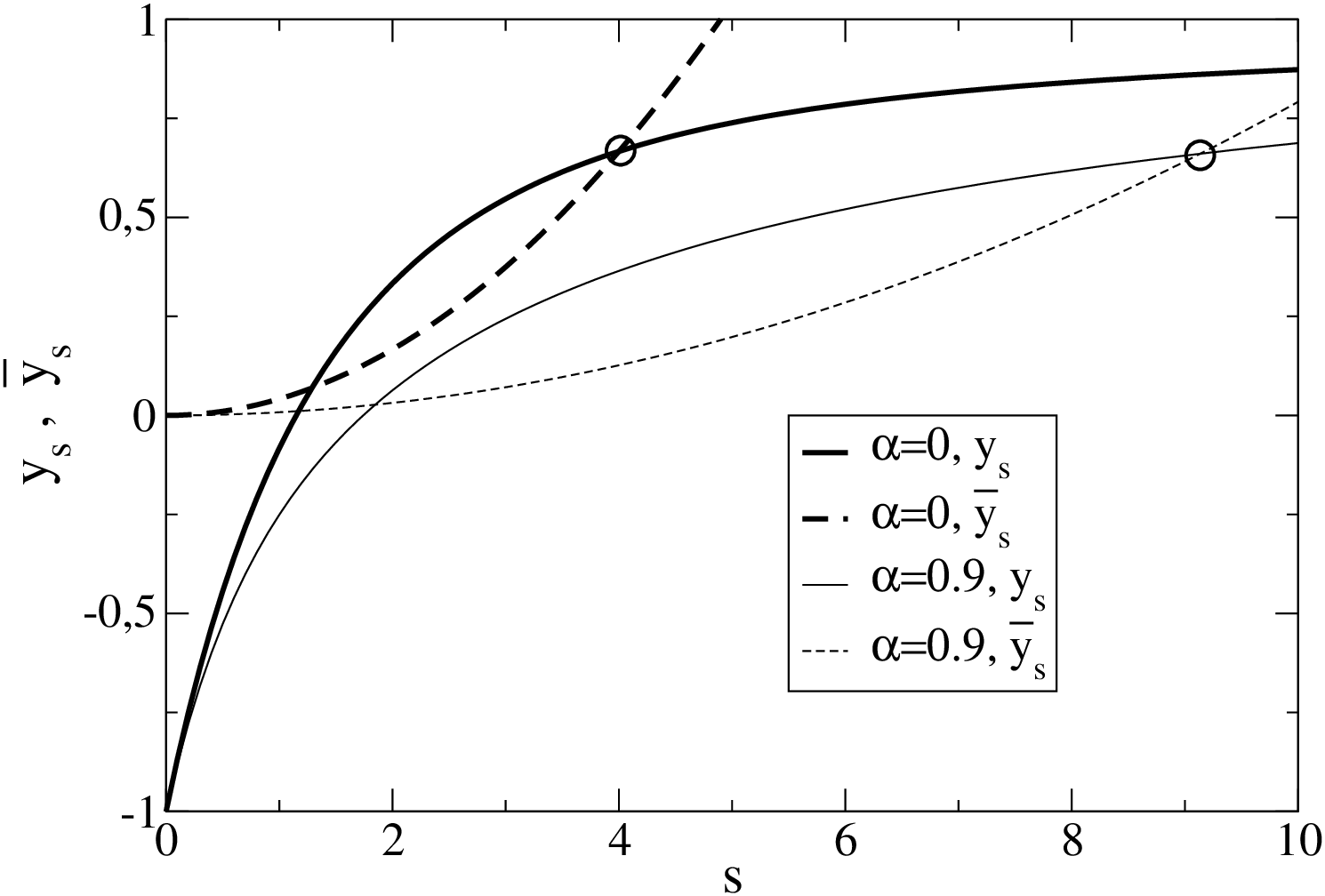}$$
\caption{Graphic solution of \Ref{64} for the marginally stable
WN-driven 2-D soft sphere model ($\sigma=1,\nu=-2$) at $\alpha=0$
(upper two curves) and at $\alpha=0.9$ (lower two curves). The
solid curves represent $y_s= \lambda_s(1) /\beta_1$ and the dashed
curves $\bar{y}_s = \lambda_2(1) s^2/(8\beta_1)= pq s^2/6$. The
largest intersection point, indicated by an open circle, gives the
exponent $a=a_d(\alpha)$ of the high energy tail $f \sim
1/c^{a+d+\nu}$.}
\Label{fig4.5}
\end{figure}

\begin{figure}
$$\includegraphics[angle=0,width=.6 \columnwidth]{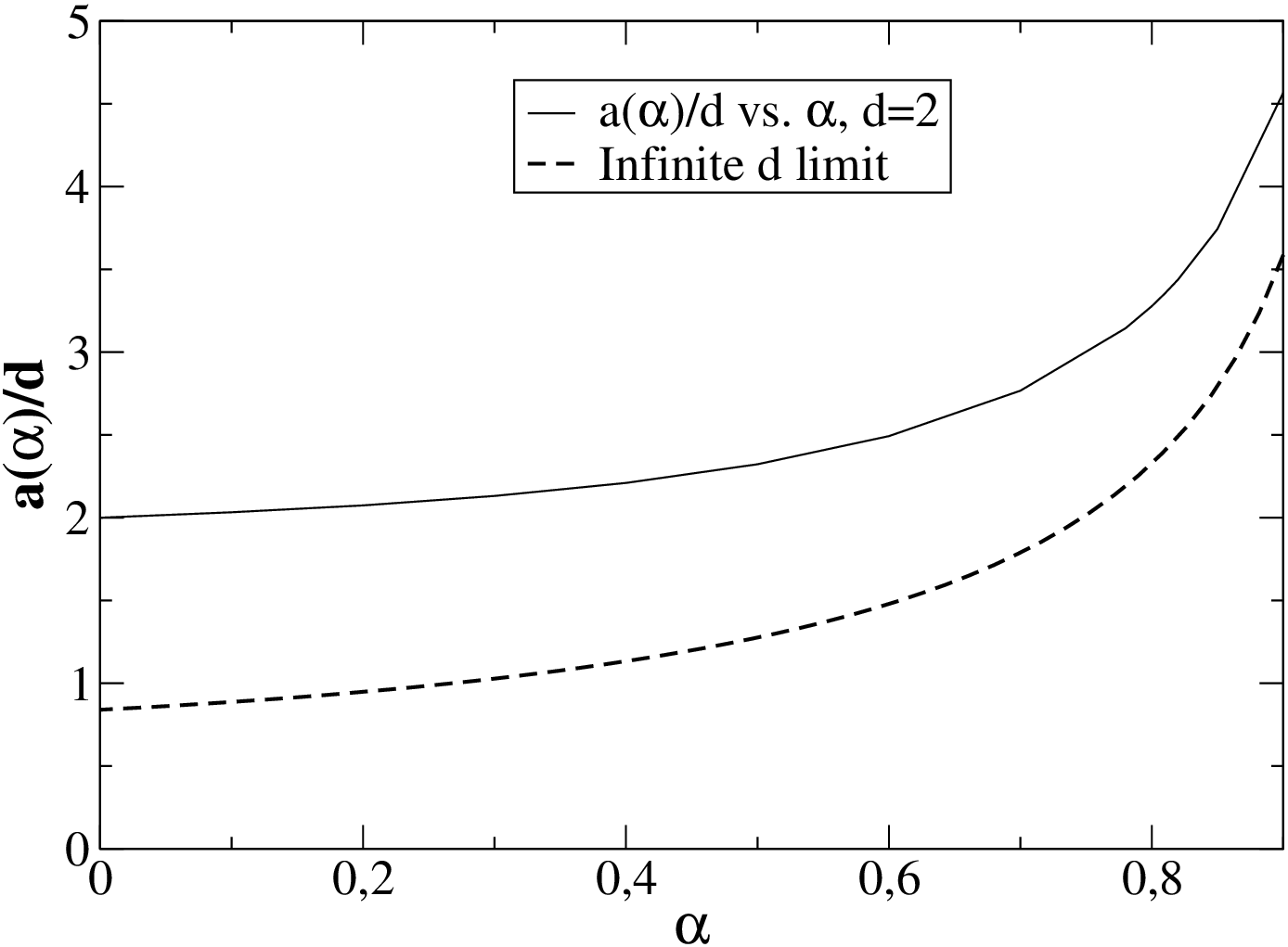}$$
\caption{ Exponent $a_d(\alpha) /d$  vs $\alpha$ of the power law
tail in the 2-D WN-driven soft sphere model $(\sigma=1,\nu=-2)$.
The solid line shows the value for $d=2$ obtained from \Ref{64}
(see also Fig.\ref{fig4.5} for the graphical solution), and the
dashed line shows the large $d-$result $x_+=a/d$, obtained from
\Ref{66}.} \Label{fig4.6}
\end{figure}

\begin{figure}
$$\includegraphics[angle=0,width=.6 \columnwidth]{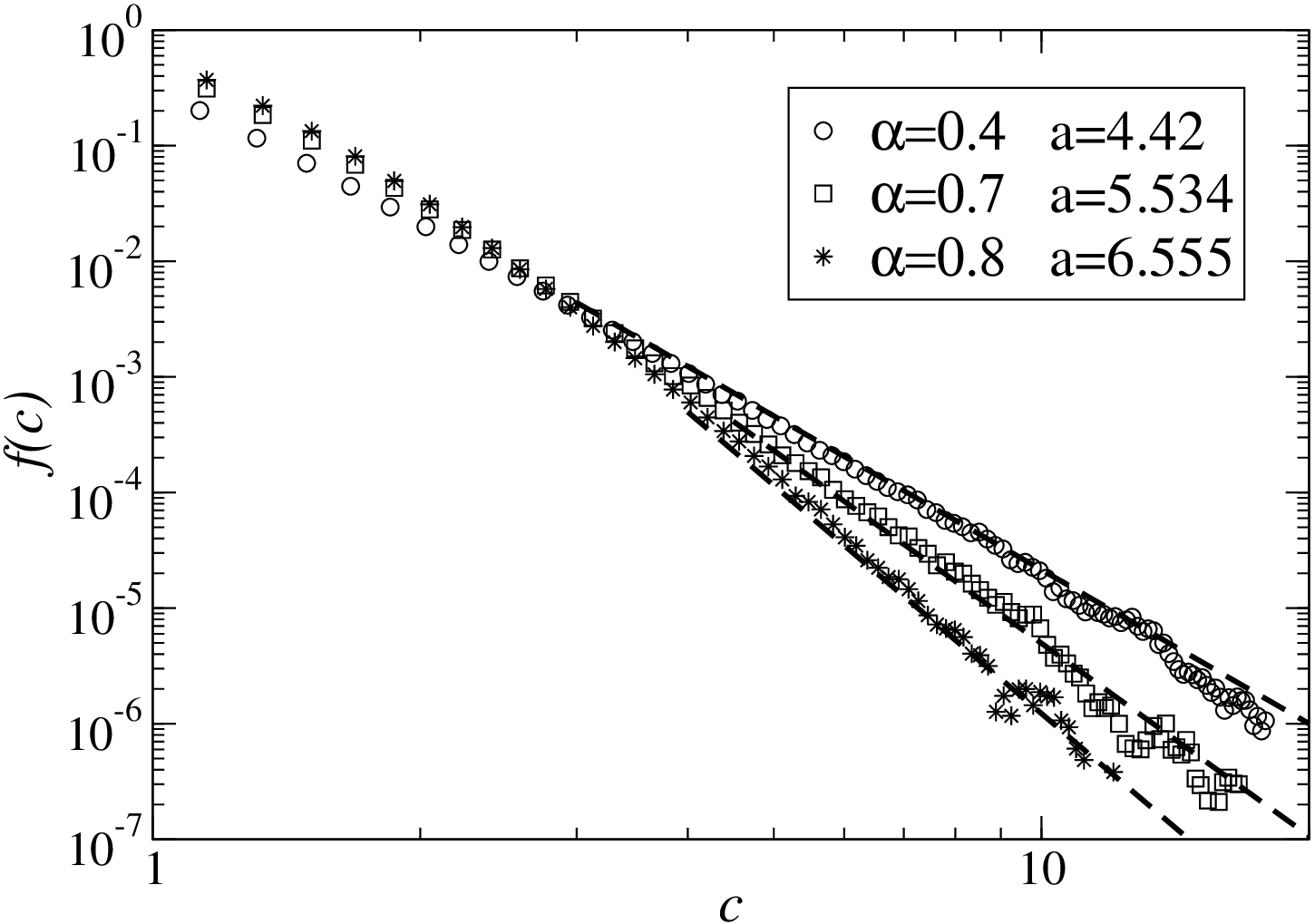}$$
\caption{ 2-D counter part of Fig.\ref{fig4.4}. Measured power law
tail of the velocity distribution in the 2-D WN-driven soft sphere
model $(\sigma=1,\nu=-2)$ at various $\alpha$. The dashed lines
correspond to the predicted power laws with exponents
$a_d(\alpha)$ given in Fig.\ref{fig4.6}. }
\Label{fig4.7}
\end{figure}

\begin{figure}
$$\includegraphics[angle=0,width=.6 \columnwidth]{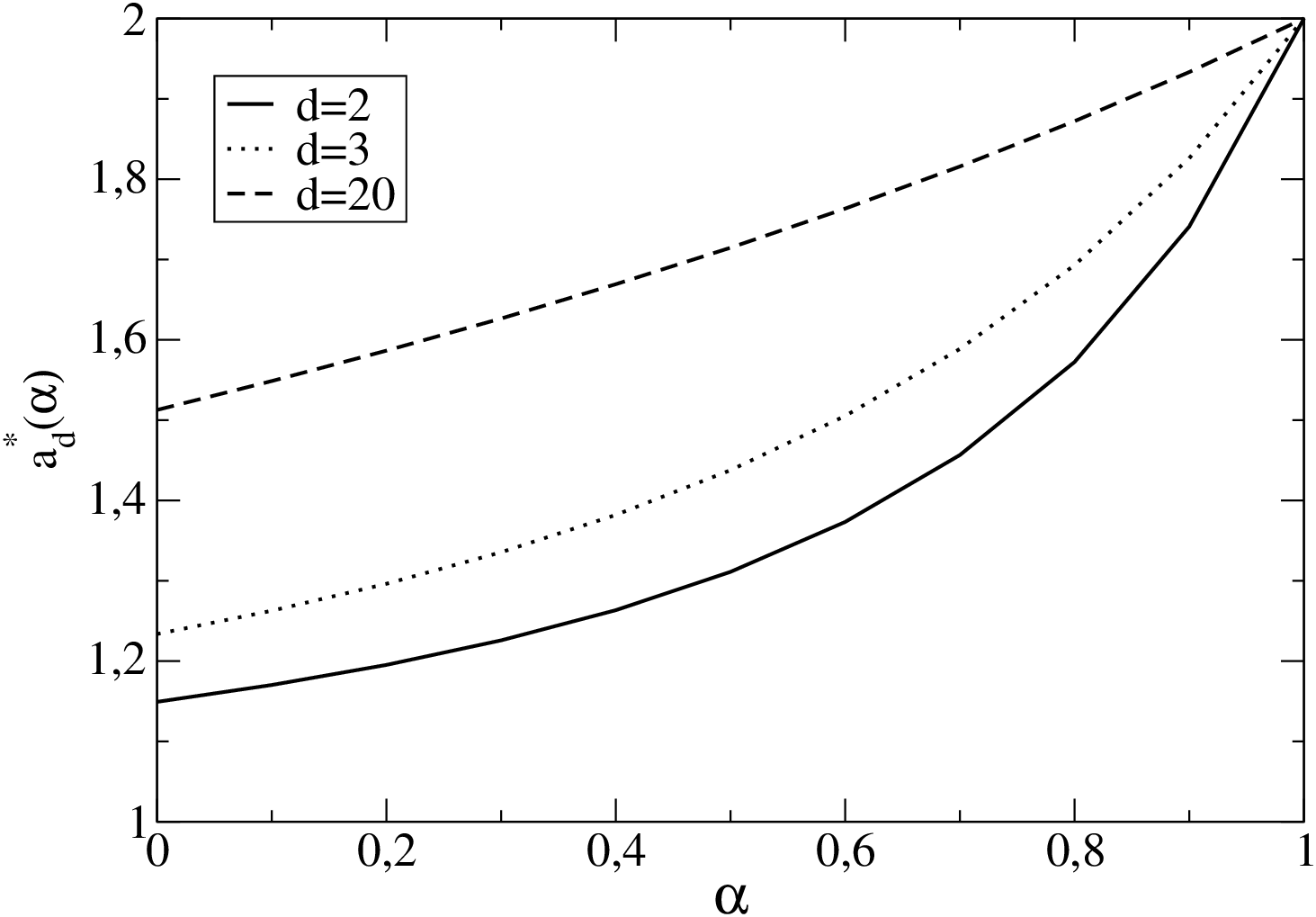}$$
\vskip 4mm
$$\includegraphics[angle=0,width=.6 \columnwidth]{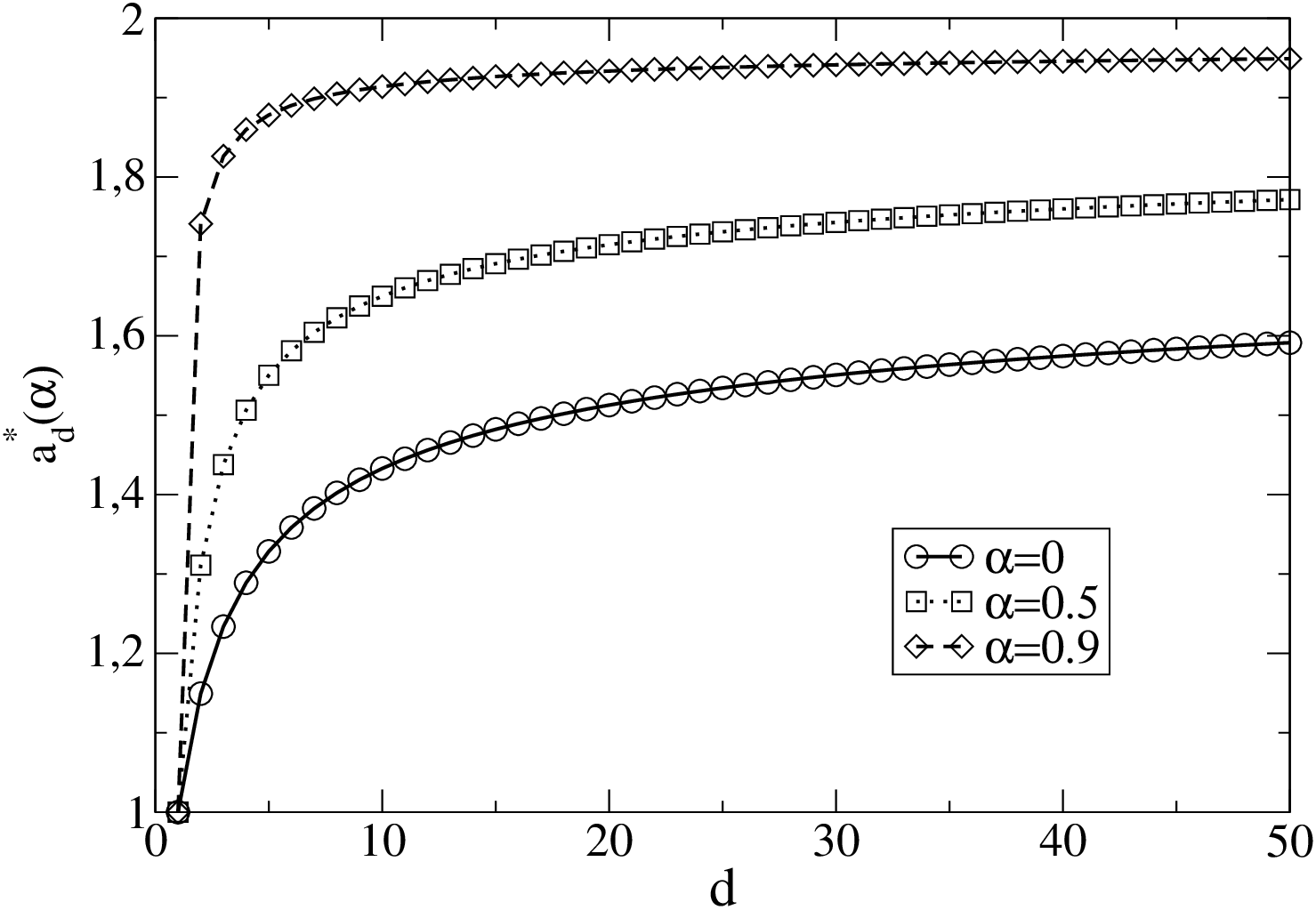}$$
\caption{Solution of $\lambda_a(\sigma)=0$ a) as a function of the
restitution coefficient $\alpha$ and b) as a function of
dimension $d$ (lower graph). The root of $\lambda_a(\sigma)=0$ is
denoted by $a^*= a^*_d(\alpha)$. }
\Label{fig5}
\end{figure}

\end{document}